\documentclass[journal=jpccck,manuscript=letter]{achemso}
\usepackage[version=3]{mhchem} % Formula subscripts using \ce{}
\usepackage{siunitx}
\DeclareSIUnit\angstrom{\text{\AA}}         %Fuer Angstrom
\author{Joshua Edzards}
\affiliation{Carl von Ossietzky Universit\"at Oldenburg, Institute of Physics, 26129 Oldenburg, Germany}
\author{Holger-Dietrich Sa{\ss}nick}
\affiliation{Carl von Ossietzky Universit\"at Oldenburg, Institute of Physics, 26129 Oldenburg, Germany}
\author{Ana Guilherme Buzanich}
\affiliation{Federal Institute for Materials Research and Testing (BAM), 12485 Berlin, Germany}
\author{Ana M. Valencia}
\affiliation{Carl von Ossietzky Universit\"at Oldenburg, Institute of Physics,  26129 Oldenburg, Germany}
\altaffiliation{Humboldt-Universit\"{a}t zu Berlin, Physics Department and IRIS Adlershof,  12489 Berlin, Germany}
\author{Franziska Emmerling}
\affiliation{Federal Institute for Materials Research and Testing (BAM), 12485 Berlin, Germany}
\author{Sebastian Beyer}
\affiliation{Federal Institute for Materials Research and Testing (BAM), 12485 Berlin, Germany}
\altaffiliation{The Chinese University of Hong Kong, Department of Biomedical Engineering, Hong Kong Special Administrative Region of China }
\email{sebastian.beyer@cuhk.edu.hk}
\author{Caterina Cocchi}
\affiliation{Carl von Ossietzky Universit\"at Oldenburg, Institute of Physics, 26129 Oldenburg, Germany}
\alsoaffiliation{Carl von Ossietzky Universit\"at Oldenburg, Center for Nanoscale Dynamics, 26129 Oldenburg, Germany}
\altaffiliation{Humboldt-Universit\"{a}t zu Berlin, Physics Department and IRIS Adlershof, 12489 Berlin, Germany}
\email{caterina.cocchi@uni-oldenburg.de}

\title{The Effects of Ligand Substituents on the Character of Zn-Coordination in Zeolitic Imidazolate Frameworks }

\begin{document}

\newpage

 \begin{abstract}
Due to their favorable properties and high porosity, zeolitic imidazolate frameworks (ZIFs) have recently received much limelight for key technologies such as energy storage, optoelectronics, sensorics, and catalysis. Despite the widespread interest in these materials, fundamental questions regarding the zinc coordination environment remain poorly understood. By focusing on zinc(II)2-methylimidazolate (ZIF-8) and its tetrahedrally coordinated analogs with Br-, Cl-, and H-substitution in the 2-ring position, we aim to clarify how variations in the local environment of Zn impact the charge distribution and the electronic properties of these materials. Our results from density-functional theory confirm the presence of a Zn coordinative bond with a large polarization that is quantitatively affected by different substituents on the organic ligand. Moreover, our findings suggest that the variations induced by the functionalization in the Zn coordination have a negligible effect on the electronic structure of the considered compounds. On the other hand, halogen terminations of the ligands lead to distinct electronic contributions in the vicinity of the frontier region which ultimately reduce the band-gap size by a few hundred meV. Experimental results obtained from X-ray absorption spectroscopy (Zn $K$-edge) confirm the trends predicted by theory and, together with them, contribute to a better understanding of the structure-property relationships that are needed to tailor ZIFs for target applications.
 \end{abstract}

\section*{TOC Graphic}
 \begin{center}
\includegraphics[height=4.45cm]{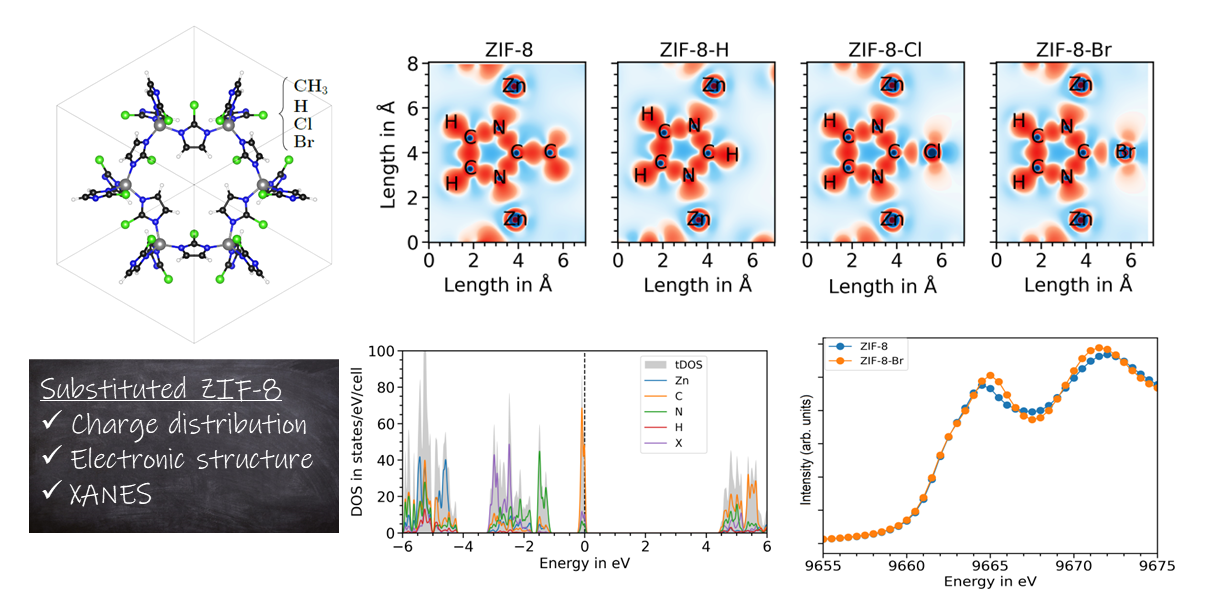}
\end{center}

\newpage

\section{Introduction}

Zeolitic imidazolate frameworks (ZIFs) are a popular family of metal-organic frameworks (MOFs) and ZIF-8, a polymorph of zinc(II)2-methylimidazolate~\cite{park+06pnas,huan+06acie}, is their most prominent member. 
Although transition-metal imidazolates have been known for decades~\cite{baum-wang64ic}, these materials have entered the spotlight after the discovery of their high porosity~\cite{yagh+95nat,li+99nat} and exceptional thermal and chemical stability~\cite{park+06pnas,huan+06acie}.
Thanks to their straightforward synthesis~\cite{beye+16cs,beye+18cis} and characterization~\cite{buza+21sm} as well as their intriguing properties~\cite{phan+10acr,tan+10pnas,chen+14jmca} even under pressure~\cite{spen+09jacs,hu+11cc,hu+13jacs,widm+19natm,choi+19jpcc,form+21jpcc} and in response to external stimuli~\cite{iaco-maur21acsami}, ZIFs have become relevant for a number of technological areas ranging from gas storage to photocatalysis, and from biomedicine to optoelectronics~\cite{esla+13cm,pime+14ChemSusChem,hoop+18amt,dai+21ccr,hu+21jcp,kneb-caro22natn,paul+22ACSOmega,tsan+23csapea}.

From a fundamental perspective, these characteristics are ruled by the electronic charge distribution in the material and by the nature of the chemical bonds therein.
Despite the general consensus regarding the ionic nature of the Zn-N bond in ZIFs~\cite{wang+15tca,yilm+19as,liu+21jcleanprod,wang+22acr}, the actual strength of this bond and the influence exerted on it by the chemical environment are still a matter of debate~\cite{bhat+18cm,widm+19natm,sark+22acie}.
In particular, recent studies have unraveled non-trivial characteristics in the electronic structure of ZIF-8~\cite{butl+17jmcc,moes+22jpcl,moes+22acsami}, which point to the existence of some degree of covalence in the bond between Zn and N. 
This hypothesis is supported by \textit{ab initio} calculations on the transport properties~\cite{butl+17jmcc} and by the vibrational activity of ZIFs~\cite{moes+22jpcl,moes+22acsami}.
Experimental evidence of the (partial) covalent character of Zn-N coordination is, to date, only indirect.
Observations of chemical bond stability of ZIF-8 and related compounds especially in acidic environments point to their susceptibility to hydrolytic cleavage, which is only known for covalent bonds~\cite{grai+03jpe}. 
Likewise, the photo-catalytic activity of ZIFs is sensitive to structural and chemical changes within the framework~\cite{du-zhou21cej}, while the polarizability of metallic ions should be hindered in a structure with purely ionic coordination.
In order to clarify the nature of Zn-N coordination in ZIF-8, a dedicated study combining both theoretical and experimental analysis is urgently needed.

Here, we present a joint investigation on ZIF-8 and related compounds with ligand substitution carried out with first-principles calculations and X-ray spectroscopy.
In this analysis, we consider halogen ligands terminations such as Cl and Br (zinc(II)2-chloroimidazole and zinc(II)2-bromoimidazole, abbreviated hereafter as ZIF-8-Cl and ZIF-8-Br), which are more electronegative than the \ce{CH3} group in zinc(II)2-methylimidazolate (ZIF-8); additionally, we include in the study the H termination in zinc(II)imidazolate (ZIF-8-H), which is expected to be slightly more electropositive than ZIF-8.
By means of density functional theory (DFT), we investigate the charge distribution of these compounds both in their (simplified) crystal structure as well as in the isolated metal-organic complex building up the framework and in the isolated imidazole unit. 
This way, we assess the role of the local chemical environment of the Zn coordination in determining the charge distribution in these systems.
Furthermore, through the analysis of the density of states, we clarify the influence of the functional groups on the character and composition of the electronic states close to the frontier.
We complement this study with the analysis of the near-edge X-ray absorption spectrum of ZIF-8 and its Br-terminated analog measured from the Zn $K$-edge.

%%%%%%%%%%%%%%%%%%%%%
\section{Systems and Methods}

\subsection{Modeling ZIF-8 and its Functionalized Analogs}

\begin{figure}
    \centering
    \includegraphics[width=0.4\textwidth]{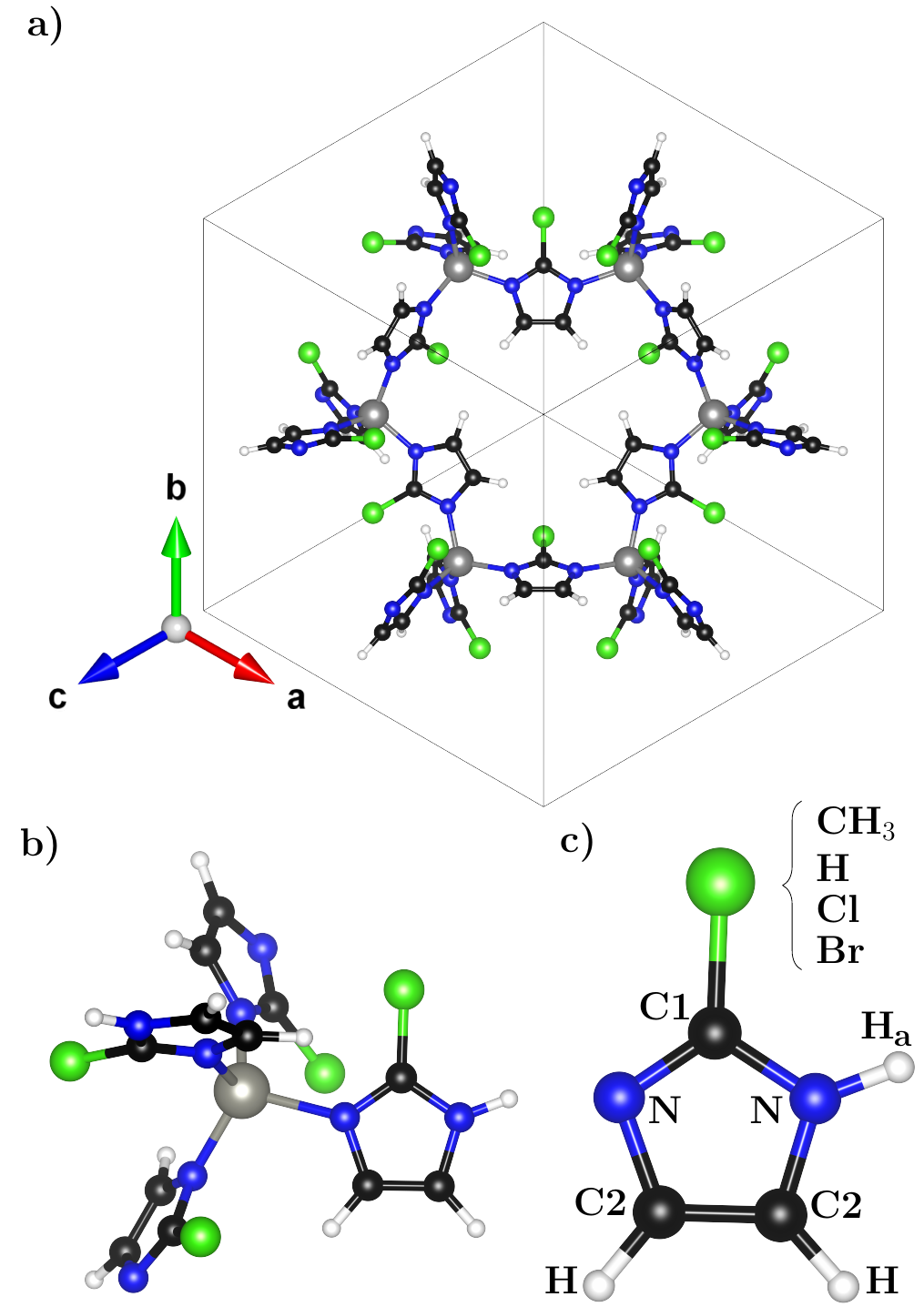}
    \caption{Ball-and-stick representation of the ZIF-8 compounds considered in this work including a) the crystal structure with \textit{R3} space-group symmetry, b) the isolated metal-organic complex replicated in the crystal, and c) the imidazole ligand. Carbon atoms are depicted in black, hydrogen atoms in white, nitrogen atoms in blue, zinc in gray, and the terminating group in green (\ce{CH3} in ZIF-8, H, Cl, and Br in the substituted siblings). The atoms in the ligand are marked in panel c): identical labels indicate equivalent atoms; H$_a$ is the proton added to the imidazole unit and to the complex to passivate one of the two N atoms.  
    }
    \label{fig:Struc_ZIF-8}
\end{figure}

In this study, we focus on ZIF-8, an established MOF structure with methyl-functionalized linker molecules.
To investigate the effects induced in the Zn coordination by different terminations of the organic ligands, we consider the ZIF-8 analogs with H-, Cl-, and Br-substituents of the methyl groups in the 2-ring position (see Figure~\ref{fig:Struc_ZIF-8}).
While the H-passivated ZIF-8 has been only predicted computationally~\cite{lewis+09rec}, the halogen-functionalized systems, have been synthesized by several groups (see, among others, Refs.~\citenum{li+09jacs,amro+11jpcc,chap+18jpcc,yagi+23pccp}) and adopted in theoretical studies as representative ZIF-8 variants~\cite{amro+11jpcc,amro+12rsca,duer+19jctc}. 
To model ZIF-8-H, we have adopted the predicted crystal structure published in Ref.~\citenum{lewis+09rec}.
On the other hand, we have taken the structural information of ZIF-8 and its halogenated counterparts directly from the experimental references.

For ZIF-8, we have taken the system with \textit{R3} space group~\cite{morris+12jpcc}.
With a script developed in-house, we have adjusted the structure provided in Ref.~\citenum{morris+12jpcc} by removing redundant H atoms.
The adopted procedure includes finding
the nearest hydrogen for each substituted C atom of the methyl group and the next H atoms with angles comprised between \SI{100}{\degree} and \SI{130}{\degree} to obtain the \ce{CH3} functional group.
For both ZIF-8-Cl and ZIF-8-Br, we took the structural data from Ref.~\citenum{chap+18jpcc}. 
In these cases, the linker molecule assumes two different orientations in the scaffold.
In our bulk models, we took the molecular orientation giving rise to a bulk with \textit{I-43m} symmetry.
All of these crystals have the same network topology (SOD)~\cite{Ke+21JCED}.  
The rhombohedral unit cell adopted in the calculations of all the considered ZIF-8-X (with X = \ce{CH3}, H, Br, Cl), see Figure~\ref{fig:Struc_ZIF-8}a, is retrieved from the initial cubic cell using the code \texttt{seek-path}~\cite{hinuma+17cms}.
To preserve the symmetry of the initial structures, the input lattice parameters of all the considered ZIF-8 variants (see Table~S1 of the Supporting Information) are kept constant during optimization, which thus entails only the minimization of the interatomic forces. 

To model the isolated complexes (Figure~\ref{fig:Struc_ZIF-8}b), the atomic positions are extracted from the ZIF-8-H crystal~\cite{lewis+09rec}. Based on this structure, the other functional groups are placed to substitute the terminating hydrogen. The same procedure is adopted for the imidazole molecule (Figure~\ref{fig:Struc_ZIF-8}c).
In these calculations, the non-periodic structures are embedded in a supercell including \SI{7.5}{\angstrom} of vacuum in each direction to prevent spurious interactions among periodic replicas.

%%%%%%%%%%%%%%%%%%%%%%%%%%%%%%%%%
\subsection{Computational Details}

All calculations presented in this work are done in the framework of DFT~\cite{hohe-kohn64pr} solving the Kohn-Sham equations~\cite{kohn-sham65pr} with the Perdew-Burke-Ernzerhof (PBE) exchange-correlation functional~\cite{pbe}. The code Quantum ESPRESSO~\cite{qe2020} is adopted with the projector augmented wave method~\cite{bloechl94prb} and pseudopotentials from the \texttt{ps-library}~\cite{DALCORSO14cms}.
A $2\times2\times2$ \textbf{k}-mesh is used to sample the Brillouin zone of all the considered periodic structures.
The plane-wave and electron density cutoffs are set to \SI{150}{Ry} and \SI{1000}{Ry}, respectively.
Structure optimization is performed by minimizing interatomic forces until the convergence threshold of 10$^{-4}$~Ry/bohr in the crystal and of 10$^{-3}$~Ry/bohr is achieved in the metal-organic complexes and in the imidazole units, respectively.
The Bader partial charges~\cite{BADER92cpl} and the planar deformation density plots have been calculated using the post-processing software package \texttt{critic2}~\cite{oter+14cpc}.
The method derived by Yu-Trinkle~\cite{yu+trin11jcp} was used to integrate the partial charges.

%%%%%%%%%%%%%%%%%%%%%%%%%%%%%%%
\subsection{Experimental Methods}
X-ray absorption near edge structure spectroscopy (XANES) measurements were performed at the BAMline (BESSY-II, Helmholtz Centre Berlin for Materials and Energy Berlin, Germany)~\cite{guil+23jcp}. The beam was monochromatized using a double-crystal monochromator (DCM) with a Si crystal with [111] orientation. The size of the beam was 3~mm (h) $\times$ 1~mm (v). The measurements were performed at Zn $K$-edge (9659~eV) in transmission, with two ionization chambers as detectors. For the pre-edge region, the energy was varied in 10~eV steps; for the region around the edge, energy was tuned first in 0.5~eV steps, and from then on in 1~eV steps. For the measurement, the samples were mixed with boron nitride, placed in polycarbonate hole plates with a thickness of 1~mm, and sealed with polyimide tape (Kapton) on both sides. Before collecting the sample spectra, a zinc foil was used as a reference for the respective K edge. The relative energies of the spectra were calibrated against the first inflection point from the first derivative of the zinc metal absorption edge. The XAS data were processed by using ATHENA which belongs to the main package IFEFFIT (v. 1.2.11)~\cite{rave-newv05jsr}. 

%%%%%%%%%%%%%%%%%%%%%%
\section{Results and Discussion}

\subsection{Structural Properties}
We start our study by inspecting the structural properties of the considered systems.
The gained information is important to support the charge-density analysis reported in the next section. 
The results obtained for the Zn-N bond in all the considered ZIFs are reported in Table~\ref{tab:avg_bond_length_Zn-N}, where the initial values in the unrelaxed input structures are included for comparison. The errors associated with the values of the optimized structures are due to the average.
The separation between Zn and N is close to 2~\AA{} in all crystal structures with variations induced by the ligand substituents.
In ZIF-8, the calculated Zn-N distance is equal to 1.989~\AA{}, slightly larger than in the initial experimental structure~\cite{morris+12jpcc} and in excellent agreement with other measurements~\cite{park+06pnas}.
Ligand substitution leads to an increase in the Zn-N bond length by less than 0.01~\AA{} upon halogen functionalization and by 0.015~\AA{} in the presence of H. 
Notice the different trend of the initial structures, where the Zn-N separation is smaller in ZIF-8-Cl compared to ZIF-8.
ZIF-8-H is the only structure in which the Zn-N distance decreases upon relaxation (see Table~\ref{tab:avg_bond_length_Zn-N}).
The other bond lengths undergo variations of the same magnitude as Zn-N in all the considered systems (see Table~S3).
In particular, it is worth mentioning that in ZIF-8 and in its halogen-functionalized counterparts, the distance between the carbon atoms C2 (see Figure~\ref{fig:Struc_ZIF-8}c) increases upon relaxation while it remains essentially equal to the experimental value in ZIF-8-H.
Likewise, the Cn-N distances (n = 1 and 2, see Figure~\ref{fig:Struc_ZIF-8}c), which are almost identical in the initial structures, rearrange themselves upon relaxation.

\begin{table}
\caption{Average Zn-N bond lengths of the initial structures and of the optimized crystals and complexes.}
\label{tab:avg_bond_length_Zn-N}
\begin{tabular}{lcccc}
    \hline
    & ZIF-8 & ZIF-8-H & ZIF-8-Cl & ZIF-8-Br \\
    \hline
    initial &
    1.965~\AA & 2.025~\AA & 1.942~\AA & 1.987~\AA \\
    crystal &
    $\SI[separate-uncertainty = true]{1.989(0.003)}{\angstrom}$ &
    $\SI[separate-uncertainty = true]{2.004(0.001)}{\angstrom}$ &
    $\SI[separate-uncertainty = true]{1.997(0.001)}{\angstrom}$ &
    $\SI[separate-uncertainty = true]{1.998(0.001)}{\angstrom}$ \\
    complex &
    $\SI[separate-uncertainty = true]{2.031(0.089)}{\angstrom}$ &
    $\SI[separate-uncertainty = true]{2.019(0.086)}{\angstrom}$ &
    $\SI[separate-uncertainty = true]{2.022(0.087)}{\angstrom}$ &
    $\SI[separate-uncertainty = true]{2.029(0.085)}{\angstrom}$ \\
    \hline
\end{tabular}
\end{table}

The same analysis performed on the isolated metal-organic complexes reveals a different trend (see Table~\ref{tab:avg_bond_length_Zn-N}).
ZIF-8 is characterized by the longest Zn-N bond (2.031~\AA{}) followed by those in the halogenated compounds (2.029~\AA{} with Br and 2.022~\AA{} with Cl).
The complex terminated with H exhibits the shortest Zn-N distance of 2.019~\AA{}.
These findings, in excellent agreement with earlier results obtained on such clusters on analogous level of theory~\cite{wang+15tca}, indicate that in the absence of any structural constraints given by the periodic boundary conditions, the ligand termination in the metal-organic complexes exerts a more pronounced influence on the local environment of the Zn coordination.
In particular, the halogen species, characterized by a significantly larger atomic radius and electronegativity than hydrogen, give rise to an increase in the Zn-N bond length, while in the bulk the opposite trend is found.
In elaborating on these values two important points should be considered.
First, the Zn-N bond lengths reported for the complexes carry an uncertainty that is almost two orders of magnitude higher than their counterparts in the crystals.
This is due to the larger variations in the non-periodic system among the four Zn-N bonds that are present therein.
Second, the atoms in the complexes are completely free to move in the DFT optimization, in contrast with those in the crystals which are constrained by the periodicity of the lattice.
In this regard, it is worth recalling that no volume optimization was performed for the crystal structures: the atoms were allowed to relax within the experimental unit cells taken as input, with the symmetry of the synthesized crystals being preserved.
Overall, the analysis of all bond lengths in the crystals and complexes reveals that only the Zn-N distances undergo significant variations (see Table~S2), thus confirming that the influence of ligand functionalization is partly counterbalanced in the periodic environment.
Our choice to study isolated metal-organic complexes as they are extracted from the crystalline scaffold, \textit{i.e.}, without any additional ``capping" of the N bonds to mimic the dative nature of the Zn-N bond in the crystal~\cite{cui-schm20jpcc}, is motivated by our interest in understanding how the Zn coordination is affected by ligand terminations in isolated and periodic arrangements.

The results obtained in this study for the considered ZIF-8 structures are in general agreement with the experimental Zn-N bond lengths reported in the literature for various organometallic compounds~\cite{hart+97jacs,bock+99jacs}. 
Specifically, in the crystal structure of hexaimidazole-zinc(II) dichloride tetrahydrate, where Zn is coordinated with N, the Zn-N distances range between 2.153~\AA{} and 2.264~\AA{}~\cite{sand-brae67}.
Shorter values are reported for dichlorobis(N-n-propylsalicylaldimine)zinc(II), where Zn is two-fold coordinated with N, and Zn-N separations are slightly below 2~\AA{} (1.998~\AA{})~\cite{torz+02polyhedron}.
This variability of the Zn-N bond length in different chemical environments reflects, on the one hand, the current discussions regarding the nature of such a bond, and, on the other hand, highlights the need for a deeper investigation. 
%

%%%%%%%%%%%%%%%%%%%%%%%%%%%%%%%%%%%%%%%%%%%%%%%%
\subsection{Partial Charge Analysis}

\begin{figure}
    \centering
    \includegraphics[width=\textwidth]{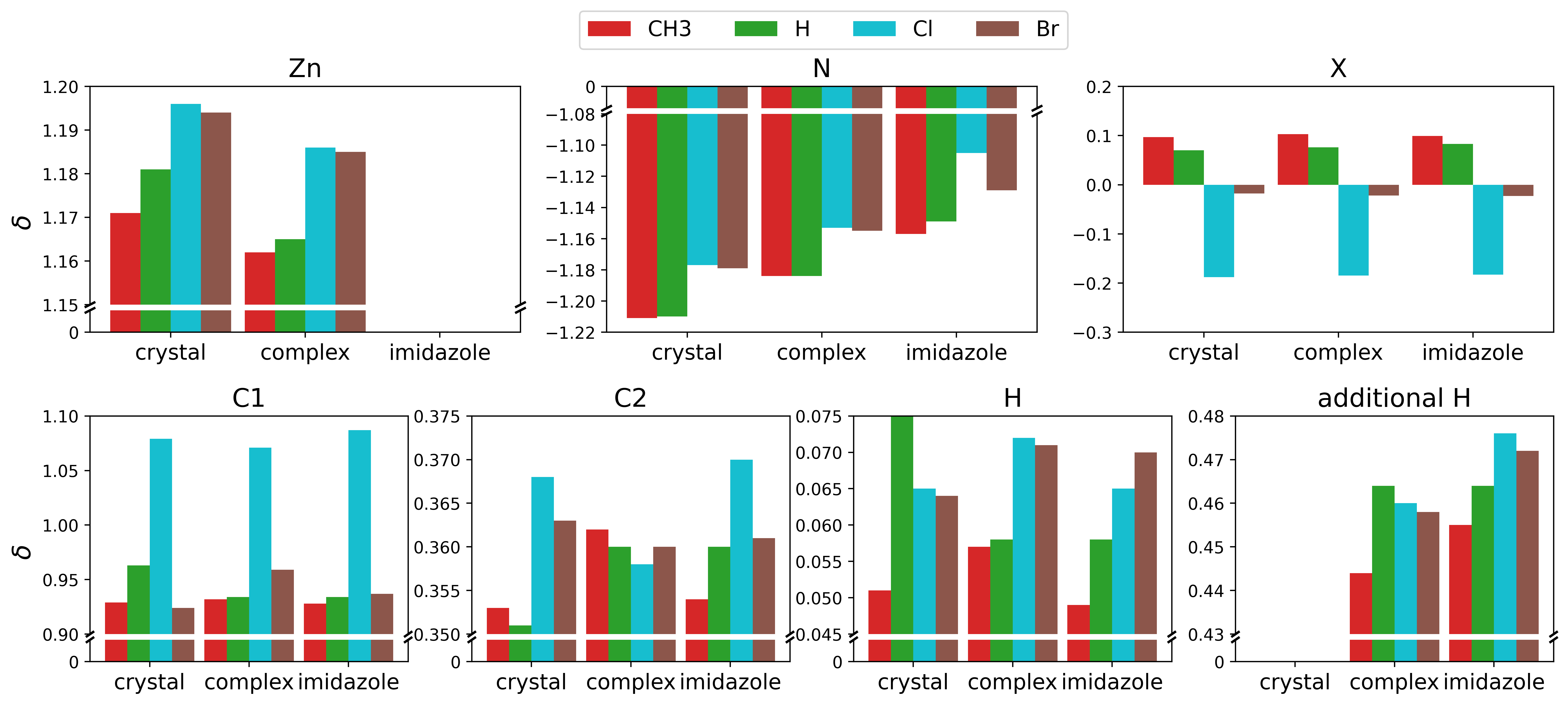}
    \caption{Partial charges per atom on Zn, N, and the functional group (X) for the considered ZIF structures. Results reported for the ``additional H'' concern only the complexes and the isolated imidazole units, where protonation of a dangling bond was needed to neutralize the system. 
    }
    \label{fig:partial_charges}
\end{figure}

Equipped with the knowledge of the structural properties presented above, we proceed with the partial-charge analysis.
This way, we aim to gain a better understanding of the electron density distribution around the Zn-N bond and of the effects of ligand functionalization.
The adopted Bader scheme~\cite{BADER92cpl}, in particular, offers reliable results for the periodic systems and the isolated clusters on equal footing.
With this method, we evaluate the relative amount of electronic charge on the atomic species of ZIF-8 and its related compounds with ligand substitution. 
The results reported in Figure~\ref{fig:partial_charges} correspond to the partial charges averaged on all atoms of a given species within the unit cell adopted in the simulations (for further details, see Supporting Information, Tables~S4-S8).
In ZIF-8, an excess of positive charge (1.17~$e^-$) is found on the metal ion while the (negative) electronic charge on nitrogen increases by -1.21~$e^-$ compared to its nominal value in the free atom.
The difference between the absolute values of these quantities is compensated by the functional group (\ce{CH3} in ZIF-8) which takes up a positive charge of 0.1~$e^-$.
The remaining charge density has a positive sign and is distributed among the C and H atoms. 
Carbon atoms bound to the functional group are labeled as C1 in Figure~\ref{fig:Struc_ZIF-8} and host a sizeable amount of charge equal to 0.93~$e^-$.
Conversely, the other two equivalent C atoms in the imidazole ring (C2) accommodate only 0.352~$e^-$ each.
The remaining $\sim$0.1~$e^-$ is found on the H atoms added for passivation (see Figure~\ref{fig:partial_charges}).
These trends can be understood by considering that C1 shares two bonds with the electronegative N atoms in contrast with C2 which only has one.

Moving now to the substituted compounds, we notice the effect of the electronegativity of the functional group in redistributing the charge density and hence in modifying the character of the Zn-N coordination.
When the \ce{CH3} group is replaced by H, the positive charge on Zn increases by 0.01~$e^-$, while the negative charge on each N atom remains almost unaltered with respect to ZIF-8 (see Figure~\ref{fig:partial_charges}). 
The difference is covered by the H atom, which donates 70$\%$ of its electronic charge to the network, and by the C2 atoms which increment their positive charge up to 0.358~$e^-$.
Upon halogen functionalization, the picture changes significantly.
In the presence of Cl termination, where Cl is the most electrophilic element among the considered functional groups, the excess of positive charge on Zn becomes almost as high as 1.20~$e^-$ while the negative charge on the N actually decreases by 0.04~$e^-$ with respect to the reference value in ZIF-8 (see Figure~\ref{fig:partial_charges}). 
This result shows that this behavior is not driven by an increased ionicity of the Zn-N coordination but is rather an effect of the substituents which modify the chemical environment of this bond. 
The Cl atom takes up -0.2~$e^-$ and induces a charge imbalance also on the C and H atoms, which all become more positively charged. 
Notice that the partial charge in C1, the C sharing a bond with Cl, is equal to 1.08~$e^-$ while on C2 it is as high as 0.368~e$^-$.
The increase of positive charge on H is a consequence of this redistribution, too (see Figure~\ref{fig:partial_charges}). 
Finally, Br termination induces similar results as Cl as far as the Zn-N bond is concerned: In this system, the metal ion hosts a positive charge of about 1.2~e$^-$ and N a negative charge of very similar magnitude (-1.18~e$^-$).
However, the lower electronegativity of Br compared to Cl leads to an almost vanishing partial charge on Br and to a charge distribution on C1 that is very close to the one in ZIF-8 and ZIF-8-H (see Figure~\ref{fig:partial_charges}).
For C2 and especially for the H atoms, the values of partial charges are close to those obtained with Cl termination but the small absolute values should be noted.

For a deeper understanding, it is interesting to evaluate the impact of the crystalline coordination in the charge distribution around the Zn-N bond.
For this purpose, we examine the results of the partial charge analysis performed on the isolated complex inside the unit cell of the considered structures (see Figure~\ref{fig:Struc_ZIF-8}b).
The absolute values of the partial charges $\delta$ on the functional groups are almost identical to those obtained for the periodic structures (see Figure~\ref{fig:partial_charges}).
The substantial similarities among the computed values of $\delta$ on C1 in the crystals and in the complexes confirm that their bonds formed with the functional groups are determined only by the charge distribution among the involved species.
In contrast, the partial charges on Zn and N are subject to substantial differences in the isolated clusters compared to the crystalline arrangements, in agreement with the variations of their distances discussed above (see Table~\ref{tab:avg_bond_length_Zn-N}).
In all complexes, regardless of the functional groups, the values of positive and negative charges on Zn and N, respectively, decrease systematically with respect to their counterparts in the crystals (see Figure~\ref{fig:partial_charges}). 
This variation can be understood by the passivation of the N dangling bonds.
The additional H atoms host a positive partial charge on the order of 0.4~$e^-$, which is maximized by the presence of the halogen terminations and minimized in ZIF-8. 

While this analysis suggests a reduction of Zn-N bond polarization in the isolated clusters compared to the crystals, the inspection of the Zn-N bond lengths computed for the complexes points to the opposite direction (see Table~\ref{tab:avg_bond_length_Zn-N}).
The Zn-N distances increase in the non-periodic systems exceeding in all cases 2~\AA{}.
In the complexes, the shorter Zn-N distance is found in the presence of H termination and the longest one with methyl. 
Despite the larger uncertainties obtained for these values compared to their counterparts in the crystals (see Table~\ref{tab:avg_bond_length_Zn-N}), these results contrast with the reduced polarization of the Zn-N coordination inferred from the partial charge distribution (Figure~\ref{fig:partial_charges}).
In an earlier DFT study on ZIF-8 complexes, similar values of the Zn-N distances are reported~\cite{wang+15tca}.
This agreement should exclude artifacts from our calculations and, concomitantly, suggests that in periodic systems, the long-range potential may contribute to enhancing the polar character of the Zn-N bond. 

Finally, we inspect the partial charge distribution in the imidazole ligands with the respective terminations.
The model systems adopted for this analysis are taken from the crystal structures and optimized \textit{in vacuo}.
While this analysis evidently does not give us any information about the Zn-N bond, since the metal ion is not part of the molecules, it enables us to better assess the charge distribution within the conjugated network and the functional group.
In the absence of Zn, the excess of negative charge on N decreases in magnitude by a few hundredths of $e^-$ in all systems (see Figure~\ref{fig:partial_charges}). 
The most significant reduction affects ZIF-8-Cl, where the electron-withdrawing ability of the halogen atom enhances the ionicity of the whole system. 
The charge distribution within the conjugated network, consisting in this case only of three C and three H atoms (see Figure~\ref{fig:Struc_ZIF-8}c) is almost identical to the one obtained in the complexes. 
From this analysis, we conclude that the presence of Zn in the complex impacts merely the local charge distribution on N.

\begin{figure}
    \centering
    \includegraphics[width=\textwidth]{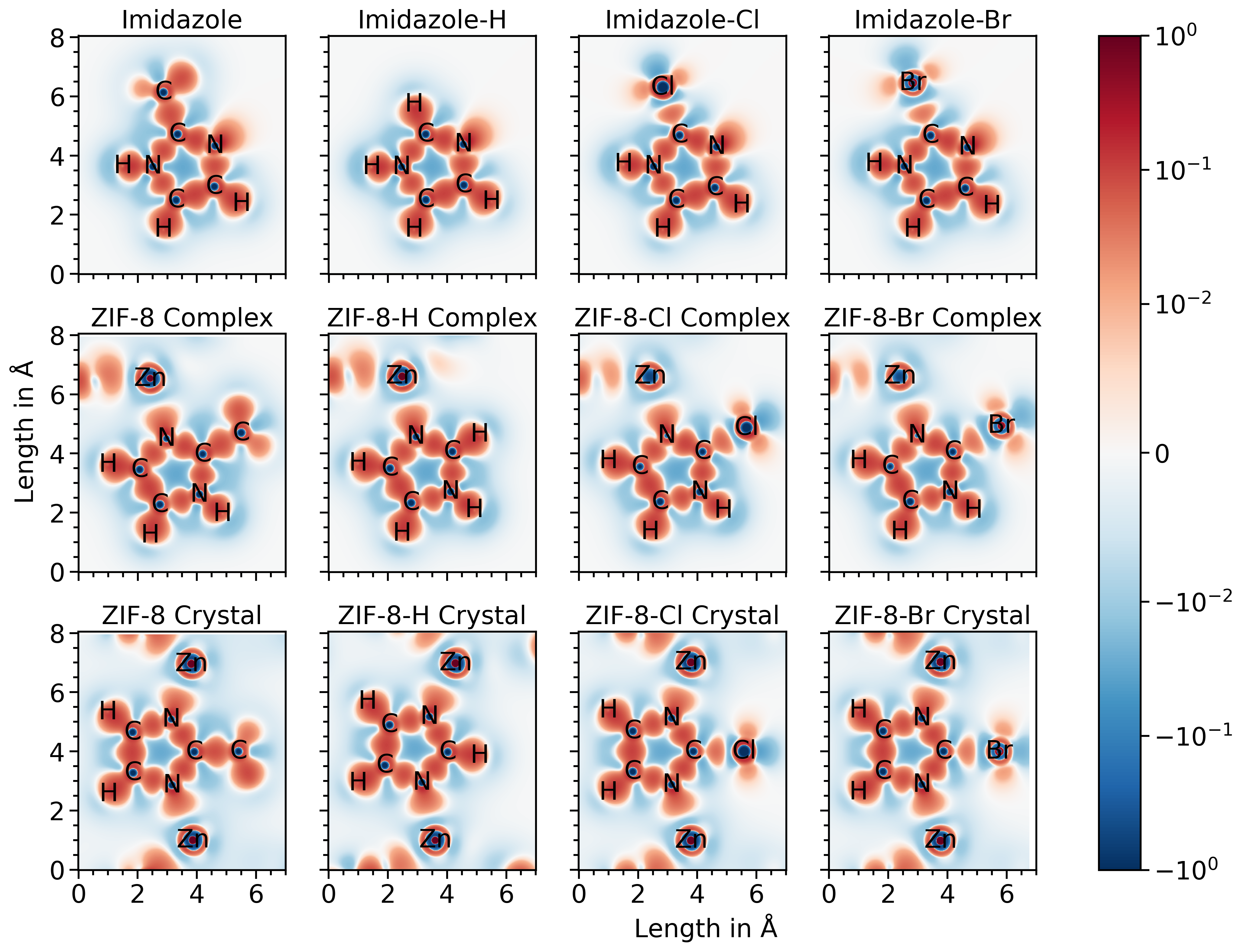}
    \caption{Electron deformation density plotted along the molecular plane of a single molecule terminated by \ce{CH3}, H, Cl, and Br, in the three considered configurations: isolated imidazole (top), complex (middle), and crystal (bottom). Electron depletion is indicated in blue and electron accumulation in red.}
    \label{fig:el_density}
\end{figure}

For a better overview of the charge distribution in the considered systems, we complement the partial charge analysis presented above with the visualization of the electron deformation density (EDD) plotted along the molecular plane of the linker molecule (see Figure~\ref{fig:el_density}).
The EDD is defined as the difference between the electron density of the considered structure and the so-called promolecular density, which is a superposition of the densities of the free atoms positioned at the molecular/crystalline sites within the structure \cite{oter+14cpc}: Positive values (red domains in Figure~\ref{fig:el_density}) represent an accumulation of electrons while negative values (blue domains in Figure~\ref{fig:el_density}) an electron depletion with respect to the free atoms.
The extension of the former over the carbon-based network can be associated with the presence of covalent bonds therein, which seems to be independent of the state of matter (crystals or complexes).
In the presence of \ce{CH3} and H ligand terminations, the functional groups are covalently bound to the backbone (Figure~\ref{fig:el_density}, left panels).
In contrast, halogen terminations are connected to the inner network by a highly polar bond, as testified by the blueish regions between them (Figure~\ref{fig:el_density}, right panels).
Likewise, the ionicity of the Zn-N bond is visible in all plots in the middle and bottom panels.

%%%%%%%%%%%%%%%%%%%%%%%%%%%%%%%%%%%%%%%%
\subsection{X-ray Absorption Spectroscopy}

\begin{figure}
    \centering
    \includegraphics[width=\textwidth]{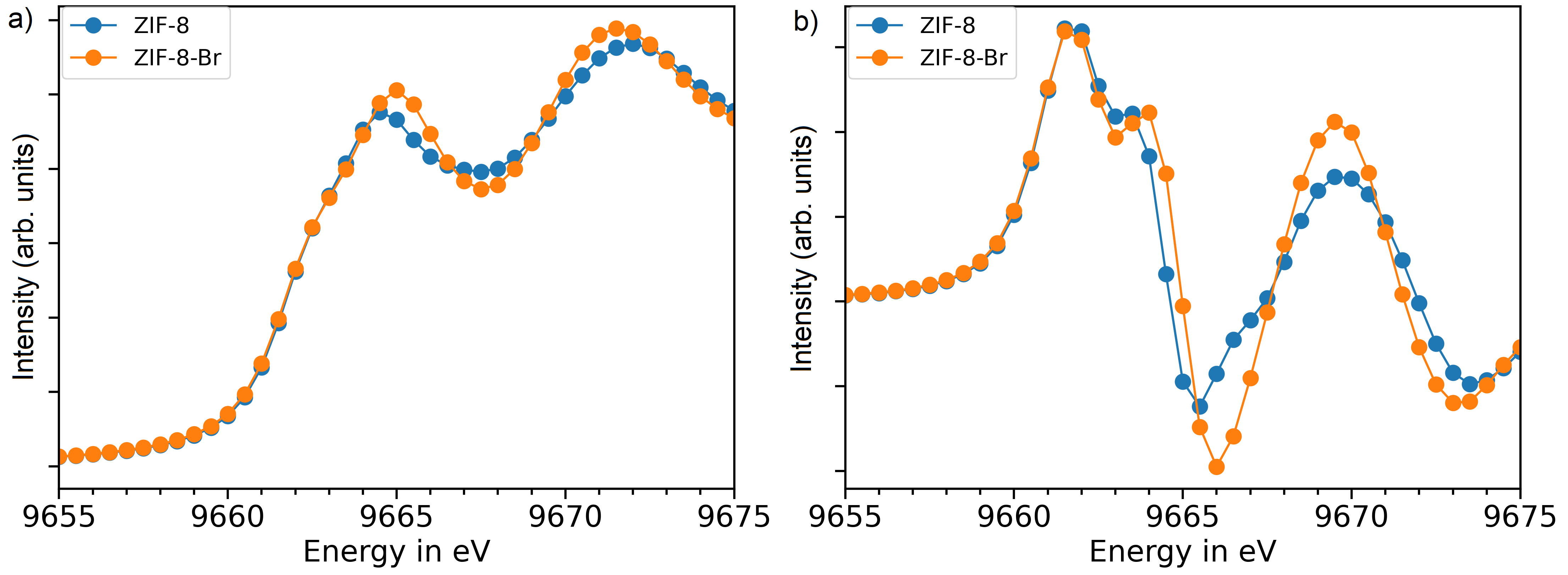}
    \caption{a) X-ray absorption from the Zn K-edge and b) its first derivative measured in zinc(II)2-bromoimidazolate and zinc(II)2-methylimidazolate.}
    \label{fig:exp-XAS}
\end{figure}

We complement the analysis of the charge-density distribution by inspecting the results of X-ray absorption measurements performed from the Zn $K$-edge.
These experiments provide an indication, at least on a qualitative level, of the Zn coordination in the ZIF-8 environment in the presence of different ligand terminations.
To this end, measurements were performed on a sample of zinc(II)2-bromoimidazolate (ZIF-8-Br) and of zinc(II)2-methylimidazolate (ZIF-8) as a reference (see Figure~\ref{fig:exp-XAS}).

Figure~\ref{fig:exp-XAS}a displays XANES spectra at the Zn $K$-edge of zinc(II)2-methylimidazolate (blue) and zinc(II)2-bromoimidazolate (orange) measured in transmission mode. The spectra reveal only subtle differences in the position of both absorption maxima, which are characteristic of (functionalized) ZIF-8 structures~\cite{boad+22jpcc}. The first maximum at 9664~eV is owed to transitions from Zn-1$s$ $\rightarrow$ Zn-4$sp$ + ligand-2$p$ states, while the second maximum at 9672 eV stems from Zn-1$s$ $\rightarrow$ Zn-4$sp$ + ligand-2$p$ orbitals.
This characteristic becomes more evident when looking at the first derivative of the signal plotted in Figure~\ref{fig:exp-XAS}b, where the spectrum of the former compound exhibits a minimum at about 0.5~eV below in energy than its counterpart in the latter.
This result can be interpreted considering that a higher-energy resonance at the onset of absorption from the Zn $K$-edge is compatible with lower electron density in this metal atom.
Hence, according to this line of reasoning, the partial charge on Zn is more positive in zinc(II)2-bromoimidazolate compared to zinc(II)2-methylimidazolate.
This is consistent with the picture provided by the DFT results.

%%%%%%%%%%%%%%%%%%%%%%%%%%%%%%%%%%%%%%%%
\subsection{Electronic Properties}

\begin{figure}
    \centering
    \includegraphics[width=\textwidth]{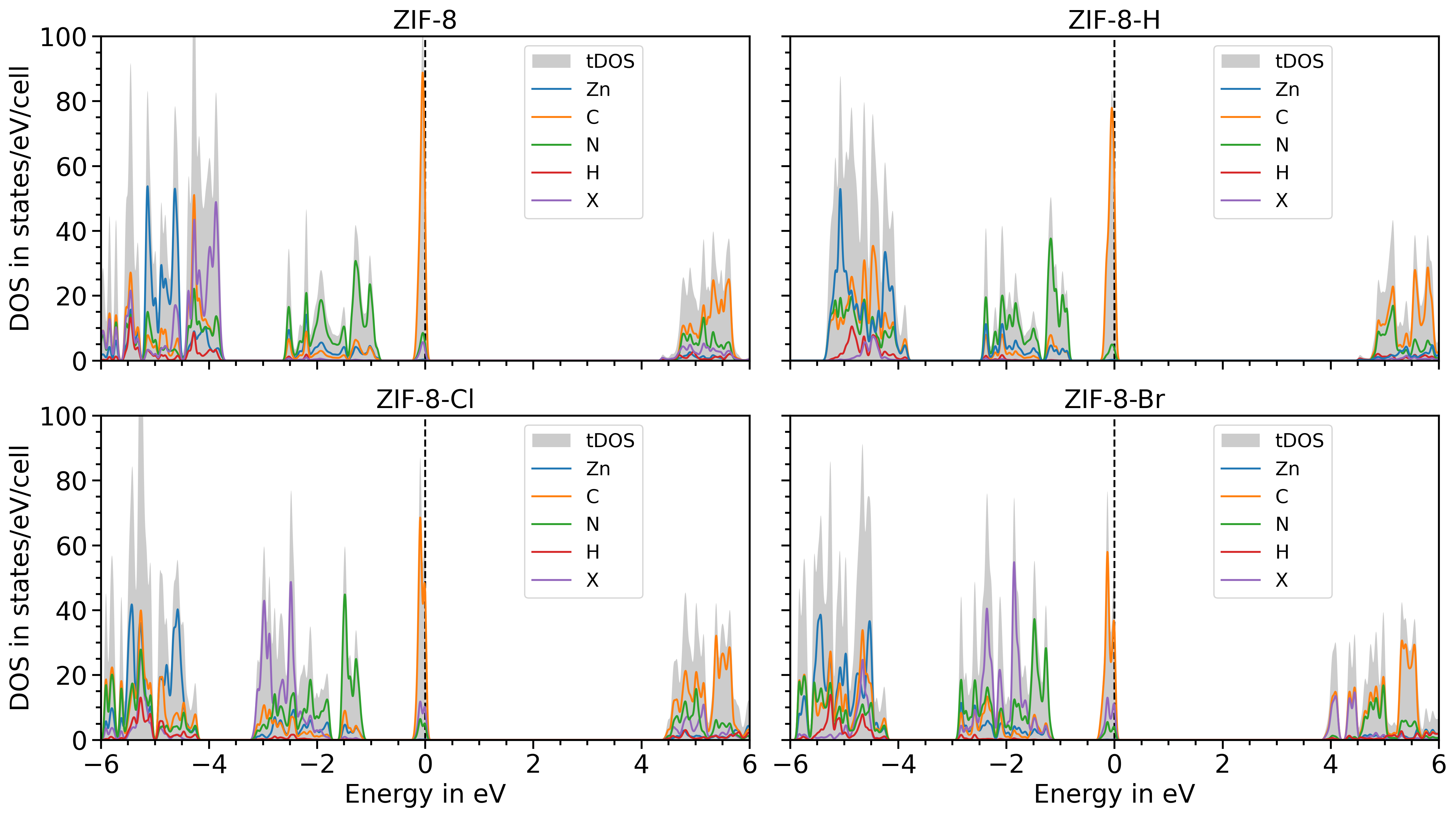}
    \caption{Projected density of states of a) ZIF-8 and its counterparts with b) H, c) Cl, and d) Br functionalization. The shaded grey area indicates the total density of states (tDOS) summing up all atomic (color code in the legend) contributions. The Fermi energy is set to 0~eV at the top of the valence band and marked by a vertical dashed bar. 
    }
    \label{fig:DOS_ZIF-8}
\end{figure}

We conclude our study by turning to the electronic structure of the considered systems.
To this end, we focus only on the periodic crystals and assess the effects of the different ligand terminations on the character and energy of the electronic states in the vicinity of the frontier.
We inspect the projected density of states (PDOS) of the considered crystal structures including the atom-resolved orbital components (see Figure~\ref{fig:DOS_ZIF-8}; in 
Figure~S1, the corresponding band structures are reported).
Before proceeding with this discussion, it is worth mentioning that these results are obtained from DFT using the PBE functional, which notoriously underestimates band gaps. 
The corresponding values in Figure~\ref{fig:DOS_ZIF-8} are therefore affected by this methodological shortcoming and should not interpreted quantitatively.
However, this issue does not undermine the following analysis of the qualitative features, which are instead well reproduced within the adopted approximations.

In all systems, the top of the valence band is dominated by a sharp peak, corresponding to a localized band manifold (see Figure~S1 and Table~S9 for the related band-gap analysis) including mainly contributions from the C atoms. 
Deeper valence states form instead a broader band with prevailing N contributions.
In the ZIF-8 with halogen terminations, Cl and Br $p$-states are present in this manifold, too (see Figure~\ref{fig:DOS_ZIF-8}c-d).
This result is unsurprising, considering the electronic structure of these atomic species with an almost filled $p$-shell at the top of their occupied region.
At even lower energies, {below -4~eV, we find the contributions of Zn $d$-orbitals hybridized with C and N states within the inner network. 
By comparing the energy distribution of the two valence-band manifolds between -3~eV and -1~eV, and below -4~eV, we can assess the influence of the various ligand terminations.
Halogen atoms enhance the energy separation of the N-dominated region shifting it down by about 0.5~eV compared to ZIF-8 and ZIF-8-H (see Figure~\ref{fig:DOS_ZIF-8}).
Moreover, the contributions of the Cl and Br atoms in the same range extend energetically this band down to about -3~eV (ZIF-8-Br, Figure~\ref{fig:DOS_ZIF-8}d) to even below that threshold (ZIF-8-Cl, Figure~\ref{fig:DOS_ZIF-8}c).
Notice, for comparison, that the N-dominated valence-band manifold extends only down to -2.5~eV in both ZIF-8 and ZIF-8-H (Figure~\ref{fig:DOS_ZIF-8}a and b).

Moving now to the conduction region, we notice at a glance that the Br-termination leads to a reduction of the band gap by more than 0.5~eV compared to ZIF-8 (see Figure~\ref{fig:DOS_ZIF-8}d and a).
The bottom of the unoccupied region in ZIF-8-Br is indeed formed by a Br-C hybridized state.
In ZIF-8-Cl, halogen contributions to the conduction bands are visible above the band onset (Figure~\ref{fig:DOS_ZIF-8}c).
Likewise, in both ZIF-8 and ZIF-8-H, the relevant contributions of the functional groups to the conduction region are above the energy range visualized in Figures~\ref{fig:DOS_ZIF-8}a and b. In these systems, the lowest-unoccupied band manifold is dominated by hybridized C and N states which are present also in the halogen-terminated structures except that in ZIF-8-Br these states appear above two unoccupied Br-based states at the conduction-band bottom and are thus up-shifted by approximately 1~eV compared to the other systems. 

The analysis of the PDOS suggests that the main role of the different ligand terminations is to alter the size of the band gap and the relative energies of the band manifold in both valence and conduction regions.
These effects are mostly pronounced with halogen functionalizations due to their electronegativity, which is known to affect the electronic properties of the carbon-based backbone hosting them through the insertion of states in the gap region (as for ZIF-8-Br here) or by downshifting the electronic states as a whole.
A similar trend was found and rationalized in the context of carbon-based nanostructures~\cite{cocc+11jpcl}.

%%%%%%%%%%%%%%%%%%%%%%%%%%%%%
\section{Summary and Conclusions}

In summary, we presented a joint first-principles and X-ray spectroscopic analysis on the influence of ligand substitution on the character of Zn-coordination in zeolitic imidazolate frameworks. We inspected the established ZIF-8 compound and its derivatives with halogen (Cl and Br) as well as H substitution in the organic ligand. By performing DFT calculations both on crystalline compounds and on their molecular building blocks as extracted from the bulk, we assessed the influence of long-range interactions in the local environment of the Zn-N coordination. 
Our results indicate that the electronegativity of the ligand substituents impacts the polarization of the Zn-N bond. The trends suggested by the DFT calculations are confirmed by the outcome of X-ray absorption measurements from the Zn $K$-edge on ZIF-8 and its Br-functionalized counterpart.
We completed our study by inspecting the electronic properties of the periodic MOFs considered in this work.
We found that the influence of the ligand substituents touches the size of the band gap, with a reduction of the order of 0.5~eV upon halogen functionalization.
Furthermore, due to their high electronegativity, Cl and Br atoms attached to the linkers lead to an overall downshift of the electronic bands in comparison to ZIF-8 or its hydrogen-terminated variant.

To conclude, this study offers new insight into the nature of Zn coordination in ZIF-8 and related compounds, and its effect on the charge-density distribution -- including the polarization of the Zn-N bond -- as well as on the electronic structure of these systems. As such, our findings confirm that chemical manipulation, for example, through ligand functionalization, represents a viable route to tune their physical properties in a flexible and controlled way.

\begin{acknowledgement}
J.E., H-D.S, A.M.V., and C.C. acknowledge funding from the German Federal Ministry of Education and Research (Professorinnenprogramm III) as well as from the State of Lower Saxony (Professorinnen f\"ur Niedersachsen and DyNano).
S.B. is supported by the Faculty of Engineering of The Chinese University of Hong Kong (CUHK), grant number 4055120.
Computational resources were provided by the North-German Supercomputing Alliance (HLRN), project nic00069, and by the local high-performance computing cluster CARL at the University of Oldenburg, financed by the German Research Foundation (Project No. INST 184/157-1 FUGG) and by the Ministry of Science and Culture of the State of Lower Saxony.
Experiments were performed at the BAMline at the BESSY-II storage ring (Helmholtz Center Berlin). We thank the Helmholtz-Zentrum Berlin für Materialien und Energie for the allocation of synchrotron radiation beamtime.
\end{acknowledgement}

\begin{suppinfo}
Additional information about the structural and electronic properties of the considered systems.
\end{suppinfo}

\section*{Data Availability Statement}
The data that support the findings of this study are openly available in
Zenodo at DOI 10.5281/zenodo.8082774.

%\bibliography{Literature}

\begin{mcitethebibliography}{66}
\providecommand*\natexlab[1]{#1}
\providecommand*\mciteSetBstSublistMode[1]{}
\providecommand*\mciteSetBstMaxWidthForm[2]{}
\providecommand*\mciteBstWouldAddEndPuncttrue
  {\def\EndOfBibitem{\unskip.}}
\providecommand*\mciteBstWouldAddEndPunctfalse
  {\let\EndOfBibitem\relax}
\providecommand*\mciteSetBstMidEndSepPunct[3]{}
\providecommand*\mciteSetBstSublistLabelBeginEnd[3]{}
\providecommand*\EndOfBibitem{}
\mciteSetBstSublistMode{f}
\mciteSetBstMaxWidthForm{subitem}{(\alph{mcitesubitemcount})}
\mciteSetBstSublistLabelBeginEnd
  {\mcitemaxwidthsubitemform\space}
  {\relax}
  {\relax}

\bibitem[Park \latin{et~al.}(2006)Park, Ni, C{\^o}t{\'e}, Choi, Huang,
  Uribe-Romo, Chae, O’Keeffe, and Yaghi]{park+06pnas}
Park,~K.~S.; Ni,~Z.; C{\^o}t{\'e},~A.~P.; Choi,~J.~Y.; Huang,~R.;
  Uribe-Romo,~F.~J.; Chae,~H.~K.; O’Keeffe,~M.; Yaghi,~O.~M. Exceptional
  chemical and thermal stability of zeolitic imidazolate frameworks.
  \emph{Proc. Natl. Acad. Sci. USA} \textbf{2006}, \emph{103},
  10186--10191\relax
\mciteBstWouldAddEndPuncttrue
\mciteSetBstMidEndSepPunct{\mcitedefaultmidpunct}
{\mcitedefaultendpunct}{\mcitedefaultseppunct}\relax
\EndOfBibitem
\bibitem[Huang \latin{et~al.}(2006)Huang, Lin, Zhang, and Chen]{huan+06acie}
Huang,~X.-C.; Lin,~Y.-Y.; Zhang,~J.-P.; Chen,~X.-M. Ligand-directed strategy
  for zeolite-type metal--organic frameworks: zinc (II) imidazolates with
  unusual zeolitic topologies. \emph{Angew.~Chem.~Int.~Ed.~} \textbf{2006},
  \emph{45}, 1557--1559\relax
\mciteBstWouldAddEndPuncttrue
\mciteSetBstMidEndSepPunct{\mcitedefaultmidpunct}
{\mcitedefaultendpunct}{\mcitedefaultseppunct}\relax
\EndOfBibitem
\bibitem[Bauman~Jr and Wang(1964)Bauman~Jr, and Wang]{baum-wang64ic}
Bauman~Jr,~J.~E.; Wang,~J.~C. Imidazole complexes of nickel (II), copper (II),
  zinc (II), and silver (I). \emph{Inorg.~Chem.~} \textbf{1964}, \emph{3},
  368--373\relax
\mciteBstWouldAddEndPuncttrue
\mciteSetBstMidEndSepPunct{\mcitedefaultmidpunct}
{\mcitedefaultendpunct}{\mcitedefaultseppunct}\relax
\EndOfBibitem
\bibitem[Yaghi \latin{et~al.}(1995)Yaghi, Li, and Li]{yagh+95nat}
Yaghi,~O.~M.; Li,~G.; Li,~H. Selective binding and removal of guests in a
  microporous metal--organic framework. \emph{Nature} \textbf{1995},
  \emph{378}, 703--706\relax
\mciteBstWouldAddEndPuncttrue
\mciteSetBstMidEndSepPunct{\mcitedefaultmidpunct}
{\mcitedefaultendpunct}{\mcitedefaultseppunct}\relax
\EndOfBibitem
\bibitem[Li \latin{et~al.}(1999)Li, Eddaoudi, O'Keeffe, and Yaghi]{li+99nat}
Li,~H.; Eddaoudi,~M.; O'Keeffe,~M.; Yaghi,~O.~M. Design and synthesis of an
  exceptionally stable and highly porous metal-organic framework. \emph{Nature}
  \textbf{1999}, \emph{402}, 276--279\relax
\mciteBstWouldAddEndPuncttrue
\mciteSetBstMidEndSepPunct{\mcitedefaultmidpunct}
{\mcitedefaultendpunct}{\mcitedefaultseppunct}\relax
\EndOfBibitem
\bibitem[Beyer \latin{et~al.}(2016)Beyer, Prinz, Sch{\"u}rmann, Feldmann,
  Zimathies, Blocki, Bald, Schneider, and Emmerling]{beye+16cs}
Beyer,~S.; Prinz,~C.; Sch{\"u}rmann,~R.; Feldmann,~I.; Zimathies,~A.;
  Blocki,~A.~M.; Bald,~I.; Schneider,~R.~J.; Emmerling,~F. Ultra-sonication of
  ZIF-67 crystals results in ZIF-67 nano-flakes. \emph{ChemistrySelect}
  \textbf{2016}, \emph{1}, 5905--5908\relax
\mciteBstWouldAddEndPuncttrue
\mciteSetBstMidEndSepPunct{\mcitedefaultmidpunct}
{\mcitedefaultendpunct}{\mcitedefaultseppunct}\relax
\EndOfBibitem
\bibitem[Beyer \latin{et~al.}(2018)Beyer, Sch{\"u}rmann, Feldmann, Blocki,
  Bald, Schneider, and Emmerling]{beye+18cis}
Beyer,~S.; Sch{\"u}rmann,~R.; Feldmann,~I.; Blocki,~A.; Bald,~I.;
  Schneider,~R.~J.; Emmerling,~F. Maintaining stable Zeolitic Imidazolate
  Framework (ZIF) templates during polyelectrolyte multilayer coating.
  \emph{Colloids Interface Sci.} \textbf{2018}, \emph{22}, 14--17\relax
\mciteBstWouldAddEndPuncttrue
\mciteSetBstMidEndSepPunct{\mcitedefaultmidpunct}
{\mcitedefaultendpunct}{\mcitedefaultseppunct}\relax
\EndOfBibitem
\bibitem[Buzanich \latin{et~al.}(2021)Buzanich, Kulow, Kabelitz, Grunewald,
  Seidel, Chapartegui-Arias, Radtke, Reinholz, Emmerling, and Beyer]{buza+21sm}
Buzanich,~A.~G.; Kulow,~A.; Kabelitz,~A.; Grunewald,~C.; Seidel,~R.;
  Chapartegui-Arias,~A.; Radtke,~M.; Reinholz,~U.; Emmerling,~F.; Beyer,~S.
  Observation of early ZIF-8 crystallization stages with X-ray absorption
  spectroscopy. \emph{Soft Matter} \textbf{2021}, \emph{17}, 331--334\relax
\mciteBstWouldAddEndPuncttrue
\mciteSetBstMidEndSepPunct{\mcitedefaultmidpunct}
{\mcitedefaultendpunct}{\mcitedefaultseppunct}\relax
\EndOfBibitem
\bibitem[Phan \latin{et~al.}(2010)Phan, Doonan, Uribe-Romo, Knobler,
  O’Keeffe, and Yaghi]{phan+10acr}
Phan,~A.; Doonan,~C.~J.; Uribe-Romo,~F.~J.; Knobler,~C.~B.; O’Keeffe,~M.;
  Yaghi,~O.~M. Synthesis, Structure, and Carbon Dioxide Capture Properties of
  Zeolitic Imidazolate Frameworks. \emph{Acc.~Chem.~Res.~} \textbf{2010},
  \emph{43}, 58--67\relax
\mciteBstWouldAddEndPuncttrue
\mciteSetBstMidEndSepPunct{\mcitedefaultmidpunct}
{\mcitedefaultendpunct}{\mcitedefaultseppunct}\relax
\EndOfBibitem
\bibitem[Tan \latin{et~al.}(2010)Tan, Bennett, and Cheetham]{tan+10pnas}
Tan,~J.~C.; Bennett,~T.~D.; Cheetham,~A.~K. Chemical structure, network
  topology, and porosity effects on the mechanical properties of Zeolitic
  Imidazolate Frameworks. \emph{Proc. Natl. Acad. Sci. USA} \textbf{2010},
  \emph{107}, 9938--9943\relax
\mciteBstWouldAddEndPuncttrue
\mciteSetBstMidEndSepPunct{\mcitedefaultmidpunct}
{\mcitedefaultendpunct}{\mcitedefaultseppunct}\relax
\EndOfBibitem
\bibitem[Chen \latin{et~al.}(2014)Chen, Yang, Zhu, and Xia]{chen+14jmca}
Chen,~B.; Yang,~Z.; Zhu,~Y.; Xia,~Y. Zeolitic imidazolate framework materials:
  recent progress in synthesis and applications. \emph{J.~Mater.~Chem.~A}
  \textbf{2014}, \emph{2}, 16811--16831\relax
\mciteBstWouldAddEndPuncttrue
\mciteSetBstMidEndSepPunct{\mcitedefaultmidpunct}
{\mcitedefaultendpunct}{\mcitedefaultseppunct}\relax
\EndOfBibitem
\bibitem[Spencer \latin{et~al.}(2009)Spencer, Angel, Ross, Hanson, and
  Howard]{spen+09jacs}
Spencer,~E.~C.; Angel,~R.~J.; Ross,~N.~L.; Hanson,~B.~E.; Howard,~J. A.~K.
  Pressure-Induced Cooperative Bond Rearrangement in a Zinc Imidazolate
  Framework: A High-Pressure Single-Crystal X-Ray Diffraction Study.
  \emph{J.~Am.~Chem.~Soc.~} \textbf{2009}, \emph{131}, 4022--4026\relax
\mciteBstWouldAddEndPuncttrue
\mciteSetBstMidEndSepPunct{\mcitedefaultmidpunct}
{\mcitedefaultendpunct}{\mcitedefaultseppunct}\relax
\EndOfBibitem
\bibitem[Hu \latin{et~al.}(2011)Hu, Kazemian, Rohani, Huang, and Song]{hu+11cc}
Hu,~Y.; Kazemian,~H.; Rohani,~S.; Huang,~Y.; Song,~Y. In situ high pressure
  study of ZIF-8 by FTIR spectroscopy. \emph{Chem.~Commun.~} \textbf{2011},
  \emph{47}, 12694--12696\relax
\mciteBstWouldAddEndPuncttrue
\mciteSetBstMidEndSepPunct{\mcitedefaultmidpunct}
{\mcitedefaultendpunct}{\mcitedefaultseppunct}\relax
\EndOfBibitem
\bibitem[Hu \latin{et~al.}(2013)Hu, Liu, Xu, Huang, and Song]{hu+13jacs}
Hu,~Y.; Liu,~Z.; Xu,~J.; Huang,~Y.; Song,~Y. Evidence of pressure enhanced CO2
  storage in ZIF-8 probed by FTIR spectroscopy. \emph{J.~Am.~Chem.~Soc.~}
  \textbf{2013}, \emph{135}, 9287--9290\relax
\mciteBstWouldAddEndPuncttrue
\mciteSetBstMidEndSepPunct{\mcitedefaultmidpunct}
{\mcitedefaultendpunct}{\mcitedefaultseppunct}\relax
\EndOfBibitem
\bibitem[Widmer \latin{et~al.}(2019)Widmer, Lampronti, Anzellini, Gaillac,
  Farsang, Zhou, Belenguer, Wilson, Palmer, Kleppe, Wharmby, Yu, Cohen, Telfer,
  Redfern, Coudert, MacLeod, and Bennett]{widm+19natm}
Widmer,~R.~N.; Lampronti,~G.~I.; Anzellini,~S.; Gaillac,~R.; Farsang,~S.;
  Zhou,~C.; Belenguer,~A.~M.; Wilson,~C.~W.; Palmer,~H.; Kleppe,~A.~K.
  \latin{et~al.}  Pressure promoted low-temperature melting of metal--organic
  frameworks. \emph{Nature~Mater.} \textbf{2019}, \emph{18}, 370--376\relax
\mciteBstWouldAddEndPuncttrue
\mciteSetBstMidEndSepPunct{\mcitedefaultmidpunct}
{\mcitedefaultendpunct}{\mcitedefaultseppunct}\relax
\EndOfBibitem
\bibitem[Choi \latin{et~al.}(2019)Choi, Im, Noh, Kim, Vogt, and
  Lee]{choi+19jpcc}
Choi,~J.; Im,~J.; Noh,~K.; Kim,~J.; Vogt,~T.; Lee,~Y. Universal Gas-Uptake
  Behavior of a Zeolitic Imidazolate Framework ZIF-8 at High Pressure.
  \emph{J.~Phys.~Chem.~C} \textbf{2019}, \emph{123}, 25769--25774\relax
\mciteBstWouldAddEndPuncttrue
\mciteSetBstMidEndSepPunct{\mcitedefaultmidpunct}
{\mcitedefaultendpunct}{\mcitedefaultseppunct}\relax
\EndOfBibitem
\bibitem[Formalik \latin{et~al.}(2021)Formalik, Mazur, Fischer, Firlej, and
  Kuchta]{form+21jpcc}
Formalik,~F.; Mazur,~B.; Fischer,~M.; Firlej,~L.; Kuchta,~B. Phonons and
  Adsorption-Induced Deformations in ZIFs: Is It Really a Gate Opening?
  \emph{J.~Phys.~Chem.~C} \textbf{2021}, \emph{125}, 7999--8005\relax
\mciteBstWouldAddEndPuncttrue
\mciteSetBstMidEndSepPunct{\mcitedefaultmidpunct}
{\mcitedefaultendpunct}{\mcitedefaultseppunct}\relax
\EndOfBibitem
\bibitem[Iacomi and Maurin(2021)Iacomi, and Maurin]{iaco-maur21acsami}
Iacomi,~P.; Maurin,~G. ResponZIF Structures: Zeolitic Imidazolate Frameworks as
  Stimuli-Responsive Materials. \emph{ACS~Appl.~Mater.~Interfaces}
  \textbf{2021}, \emph{13}, 50602--50642\relax
\mciteBstWouldAddEndPuncttrue
\mciteSetBstMidEndSepPunct{\mcitedefaultmidpunct}
{\mcitedefaultendpunct}{\mcitedefaultseppunct}\relax
\EndOfBibitem
\bibitem[Eslava \latin{et~al.}(2013)Eslava, Zhang, Esconjauregui, Yang,
  Vanstreels, Baklanov, and Saiz]{esla+13cm}
Eslava,~S.; Zhang,~L.; Esconjauregui,~S.; Yang,~J.; Vanstreels,~K.;
  Baklanov,~M.~R.; Saiz,~E. Metal-organic framework ZIF-8 films as low-$\kappa$
  dielectrics in microelectronics. \emph{Chem.~Mater.~} \textbf{2013},
  \emph{25}, 27--33\relax
\mciteBstWouldAddEndPuncttrue
\mciteSetBstMidEndSepPunct{\mcitedefaultmidpunct}
{\mcitedefaultendpunct}{\mcitedefaultseppunct}\relax
\EndOfBibitem
\bibitem[Pimentel \latin{et~al.}(2014)Pimentel, Parulkar, Zhou, Brunelli, and
  Lively]{pime+14ChemSusChem}
Pimentel,~B.~R.; Parulkar,~A.; Zhou,~E.-k.; Brunelli,~N.~A.; Lively,~R.~P.
  Zeolitic imidazolate frameworks: next-generation materials for
  energy-efficient gas separations. \emph{ChemSusChem} \textbf{2014}, \emph{7},
  3202--3240\relax
\mciteBstWouldAddEndPuncttrue
\mciteSetBstMidEndSepPunct{\mcitedefaultmidpunct}
{\mcitedefaultendpunct}{\mcitedefaultseppunct}\relax
\EndOfBibitem
\bibitem[Hoop \latin{et~al.}(2018)Hoop, Walde, Ricc{\`o}, Mushtaq, Terzopoulou,
  Chen, deMello, Doonan, Falcaro, Nelson, Puigmarti, and Panea]{hoop+18amt}
Hoop,~M.; Walde,~C.~F.; Ricc{\`o},~R.; Mushtaq,~F.; Terzopoulou,~A.;
  Chen,~X.-Z.; deMello,~A.~J.; Doonan,~C.~J.; Falcaro,~P.; Nelson,~B.~J.
  \latin{et~al.}  Biocompatibility characteristics of the metal organic
  framework ZIF-8 for therapeutical applications. \emph{Appl.~Mater.~Today}
  \textbf{2018}, \emph{11}, 13--21\relax
\mciteBstWouldAddEndPuncttrue
\mciteSetBstMidEndSepPunct{\mcitedefaultmidpunct}
{\mcitedefaultendpunct}{\mcitedefaultseppunct}\relax
\EndOfBibitem
\bibitem[Dai \latin{et~al.}(2021)Dai, Yuan, Jiang, Wang, Zhang, Zhang, and
  Xiong]{dai+21ccr}
Dai,~H.; Yuan,~X.; Jiang,~L.; Wang,~H.; Zhang,~J.; Zhang,~J.; Xiong,~T. Recent
  advances on ZIF-8 composites for adsorption and photocatalytic wastewater
  pollutant removal: Fabrication, applications and perspective.
  \emph{Coord.~Chem.~Rev.~} \textbf{2021}, \emph{441}, 213985\relax
\mciteBstWouldAddEndPuncttrue
\mciteSetBstMidEndSepPunct{\mcitedefaultmidpunct}
{\mcitedefaultendpunct}{\mcitedefaultseppunct}\relax
\EndOfBibitem
\bibitem[Hu \latin{et~al.}(2021)Hu, Pattengale, and Huang]{hu+21jcp}
Hu,~W.; Pattengale,~B.; Huang,~J. Zeolitic imidazolate frameworks as intrinsic
  light harvesting and charge separation materials for photocatalysis.
  \emph{J.~Chem.~Phys.~} \textbf{2021}, \emph{154}, 240901\relax
\mciteBstWouldAddEndPuncttrue
\mciteSetBstMidEndSepPunct{\mcitedefaultmidpunct}
{\mcitedefaultendpunct}{\mcitedefaultseppunct}\relax
\EndOfBibitem
\bibitem[Knebel and Caro(2022)Knebel, and Caro]{kneb-caro22natn}
Knebel,~A.; Caro,~J. Metal--organic frameworks and covalent organic frameworks
  as disruptive membrane materials for energy-efficient gas separation.
  \emph{Nature~Nanotechnol.} \textbf{2022}, \emph{17}, 911--923\relax
\mciteBstWouldAddEndPuncttrue
\mciteSetBstMidEndSepPunct{\mcitedefaultmidpunct}
{\mcitedefaultendpunct}{\mcitedefaultseppunct}\relax
\EndOfBibitem
\bibitem[Paul \latin{et~al.}(2022)Paul, Banga, Muthukumar, and
  Prasad]{paul+22ACSOmega}
Paul,~A.; Banga,~I.~K.; Muthukumar,~S.; Prasad,~S. Engineering the ZIF-8 Pore
  for Electrochemical Sensor Applications - A Mini Review. \emph{ACS~Omega}
  \textbf{2022}, \emph{7}, 26993--27003\relax
\mciteBstWouldAddEndPuncttrue
\mciteSetBstMidEndSepPunct{\mcitedefaultmidpunct}
{\mcitedefaultendpunct}{\mcitedefaultseppunct}\relax
\EndOfBibitem
\bibitem[Tsang \latin{et~al.}(2023)Tsang, Cheung, and Beyer]{tsan+23csapea}
Tsang,~C.~Y.; Cheung,~M. C.~Y.; Beyer,~S. Assessing the colloidal stability of
  copper doped ZIF-8 in water and serum. \emph{Colloids Surf. A Physicochem.
  Eng. Asp} \textbf{2023}, \emph{656}, 130452\relax
\mciteBstWouldAddEndPuncttrue
\mciteSetBstMidEndSepPunct{\mcitedefaultmidpunct}
{\mcitedefaultendpunct}{\mcitedefaultseppunct}\relax
\EndOfBibitem
\bibitem[Wang \latin{et~al.}(2015)Wang, Zhao, Xu, Wang, Ding, Lu, and
  Guo]{wang+15tca}
Wang,~H.; Zhao,~L.; Xu,~W.; Wang,~S.; Ding,~Q.; Lu,~X.; Guo,~W. The properties
  of the bonding between CO and ZIF-8 structures: a density functional theory
  study. \emph{Theor.~Chem.~Acta} \textbf{2015}, \emph{134}, 1--9\relax
\mciteBstWouldAddEndPuncttrue
\mciteSetBstMidEndSepPunct{\mcitedefaultmidpunct}
{\mcitedefaultendpunct}{\mcitedefaultseppunct}\relax
\EndOfBibitem
\bibitem[Yilmaz \latin{et~al.}(2019)Yilmaz, Peh, Zhao, and Ho]{yilm+19as}
Yilmaz,~G.; Peh,~S.~B.; Zhao,~D.; Ho,~G.~W. Atomic-and Molecular-Level Design
  of Functional Metal--Organic Frameworks (MOFs) and Derivatives for Energy and
  Environmental Applications. \emph{Adv.~Sci.} \textbf{2019}, \emph{6},
  1901129\relax
\mciteBstWouldAddEndPuncttrue
\mciteSetBstMidEndSepPunct{\mcitedefaultmidpunct}
{\mcitedefaultendpunct}{\mcitedefaultseppunct}\relax
\EndOfBibitem
\bibitem[Liu \latin{et~al.}(2021)Liu, Huo, Wang, Yu, Ai, Chen, Zhang, Chen,
  Song, Alharbi, \latin{et~al.} others]{liu+21jcleanprod}
Liu,~Y.; Huo,~Y.; Wang,~X.; Yu,~S.; Ai,~Y.; Chen,~Z.; Zhang,~P.; Chen,~L.;
  Song,~G.; Alharbi,~N.~S. \latin{et~al.}  Impact of metal ions and organic
  ligands on uranium removal properties by zeolitic imidazolate framework
  materials. \emph{J. Clean. Prod.} \textbf{2021}, \emph{278}, 123216\relax
\mciteBstWouldAddEndPuncttrue
\mciteSetBstMidEndSepPunct{\mcitedefaultmidpunct}
{\mcitedefaultendpunct}{\mcitedefaultseppunct}\relax
\EndOfBibitem
\bibitem[Wang \latin{et~al.}(2022)Wang, Pei, Kalmutzki, Yang, and
  Yaghi]{wang+22acr}
Wang,~H.; Pei,~X.; Kalmutzki,~M.~J.; Yang,~J.; Yaghi,~O.~M. Large Cages of
  Zeolitic Imidazolate Frameworks. \emph{Acc.~Chem.~Res.~} \textbf{2022},
  \emph{55}, 707--721\relax
\mciteBstWouldAddEndPuncttrue
\mciteSetBstMidEndSepPunct{\mcitedefaultmidpunct}
{\mcitedefaultendpunct}{\mcitedefaultseppunct}\relax
\EndOfBibitem
\bibitem[Bhattacharyya \latin{et~al.}(2018)Bhattacharyya, Han, Kim, Chiang,
  Jayachandrababu, Hungerford, Dutzer, Ma, Walton, Sholl, and Nair]{bhat+18cm}
Bhattacharyya,~S.; Han,~R.; Kim,~W.-G.; Chiang,~Y.; Jayachandrababu,~K.~C.;
  Hungerford,~J.~T.; Dutzer,~M.~R.; Ma,~C.; Walton,~K.~S.; Sholl,~D.~S.
  \latin{et~al.}  Acid Gas Stability of Zeolitic Imidazolate Frameworks:
  Generalized Kinetic and Thermodynamic Characteristics. \emph{Chem.~Mater.~}
  \textbf{2018}, \emph{30}, 4089--4101\relax
\mciteBstWouldAddEndPuncttrue
\mciteSetBstMidEndSepPunct{\mcitedefaultmidpunct}
{\mcitedefaultendpunct}{\mcitedefaultseppunct}\relax
\EndOfBibitem
\bibitem[Sarkar \latin{et~al.}(2022)Sarkar, Gr{\o}nbech, Mamakhel, Bondesgaard,
  Sugimoto, Nishibori, and Iversen]{sark+22acie}
Sarkar,~S.; Gr{\o}nbech,~T. B.~E.; Mamakhel,~A.; Bondesgaard,~M.; Sugimoto,~K.;
  Nishibori,~E.; Iversen,~B.~B. X-ray Electron Density Study of the Chemical
  Bonding Origin of Glass Formation in Metal--Organic Frameworks.
  \emph{Angew.~Chem.~Int.~Ed.~} \textbf{2022}, e202202742\relax
\mciteBstWouldAddEndPuncttrue
\mciteSetBstMidEndSepPunct{\mcitedefaultmidpunct}
{\mcitedefaultendpunct}{\mcitedefaultseppunct}\relax
\EndOfBibitem
\bibitem[Butler \latin{et~al.}(2017)Butler, Worrall, Molloy, Hendon, Attfield,
  Dryfe, and Walsh]{butl+17jmcc}
Butler,~K.~T.; Worrall,~S.~D.; Molloy,~C.~D.; Hendon,~C.~H.; Attfield,~M.~P.;
  Dryfe,~R.~A.; Walsh,~A. Electronic structure design for nanoporous,
  electrically conductive zeolitic imidazolate frameworks.
  \emph{J.~Mater.~Chem.~C} \textbf{2017}, \emph{5}, 7726--7731\relax
\mciteBstWouldAddEndPuncttrue
\mciteSetBstMidEndSepPunct{\mcitedefaultmidpunct}
{\mcitedefaultendpunct}{\mcitedefaultseppunct}\relax
\EndOfBibitem
\bibitem[M\"oslein and Tan(2022)M\"oslein, and Tan]{moes+22jpcl}
M\"oslein,~A.~F.; Tan,~J.-C. Vibrational Modes and Terahertz Phenomena of the
  Large-Cage Zeolitic Imidazolate Framework-71. \emph{J.~Phys.~Chem.~Lett.}
  \textbf{2022}, \emph{13}, 2838--2844\relax
\mciteBstWouldAddEndPuncttrue
\mciteSetBstMidEndSepPunct{\mcitedefaultmidpunct}
{\mcitedefaultendpunct}{\mcitedefaultseppunct}\relax
\EndOfBibitem
\bibitem[M\"oslein \latin{et~al.}(2022)M\"oslein, Don{\`a}, Civalleri, and
  Tan]{moes+22acsami}
M\"oslein,~A.~F.; Don{\`a},~L.; Civalleri,~B.; Tan,~J.-C. Defect Engineering in
  Metal--Organic Framework Nanocrystals: Implications for Mechanical Properties
  and Performance. \emph{ACS~Appl.~Mater.~Interfaces} \textbf{2022}, \emph{5},
  6398--6409\relax
\mciteBstWouldAddEndPuncttrue
\mciteSetBstMidEndSepPunct{\mcitedefaultmidpunct}
{\mcitedefaultendpunct}{\mcitedefaultseppunct}\relax
\EndOfBibitem
\bibitem[Graiver \latin{et~al.}(2003)Graiver, Farminer, and
  Narayan]{grai+03jpe}
Graiver,~D.; Farminer,~K.; Narayan,~R. A review of the fate and effects of
  silicones in the environment. \emph{J.~Polym.~Environ.} \textbf{2003},
  \emph{11}, 129--136\relax
\mciteBstWouldAddEndPuncttrue
\mciteSetBstMidEndSepPunct{\mcitedefaultmidpunct}
{\mcitedefaultendpunct}{\mcitedefaultseppunct}\relax
\EndOfBibitem
\bibitem[Du and Zhou(2021)Du, and Zhou]{du-zhou21cej}
Du,~X.; Zhou,~M. Strategies to enhance catalytic performance of metal-organic
  frameworks in sulfate radical-based advanced oxidation processes for organic
  pollutants removal. \emph{Chem.~Eng.~J.} \textbf{2021}, \emph{403},
  126346\relax
\mciteBstWouldAddEndPuncttrue
\mciteSetBstMidEndSepPunct{\mcitedefaultmidpunct}
{\mcitedefaultendpunct}{\mcitedefaultseppunct}\relax
\EndOfBibitem
\bibitem[Lewis \latin{et~al.}(2009)Lewis, Ruiz-Salvador, Gómez,
  Rodriguez-Albelo, Coudert, Slater, Cheetham, and
  Mellot-Draznieks]{lewis+09rec}
Lewis,~D.~W.; Ruiz-Salvador,~A.~R.; Gómez,~A.; Rodriguez-Albelo,~L.~M.;
  Coudert,~F.-X.; Slater,~B.; Cheetham,~A.~K.; Mellot-Draznieks,~C. Zeolitic
  imidazole frameworks: structural and energetics trends compared with their
  zeolite analogues. \emph{CrystEngComm} \textbf{2009}, \emph{11},
  2272--2276\relax
\mciteBstWouldAddEndPuncttrue
\mciteSetBstMidEndSepPunct{\mcitedefaultmidpunct}
{\mcitedefaultendpunct}{\mcitedefaultseppunct}\relax
\EndOfBibitem
\bibitem[Li \latin{et~al.}(2009)Li, Olson, Seidel, Emge, Gong, Zeng, and
  Li]{li+09jacs}
Li,~K.; Olson,~D.~H.; Seidel,~J.; Emge,~T.~J.; Gong,~H.; Zeng,~H.; Li,~J.
  Zeolitic Imidazolate Frameworks for Kinetic Separation of Propane and
  Propene. \emph{J.~Am.~Chem.~Soc.~} \textbf{2009}, \emph{131},
  10368--10369\relax
\mciteBstWouldAddEndPuncttrue
\mciteSetBstMidEndSepPunct{\mcitedefaultmidpunct}
{\mcitedefaultendpunct}{\mcitedefaultseppunct}\relax
\EndOfBibitem
\bibitem[Amrouche \latin{et~al.}(2011)Amrouche, Aguado, Pérez-Pellitero,
  Chizallet, Siperstein, Farrusseng, Bats, and Nieto-Draghi]{amro+11jpcc}
Amrouche,~H.; Aguado,~S.; Pérez-Pellitero,~J.; Chizallet,~C.; Siperstein,~F.;
  Farrusseng,~D.; Bats,~N.; Nieto-Draghi,~C. Experimental and Computational
  Study of Functionality Impact on Sodalite–Zeolitic Imidazolate Frameworks
  for CO2 Separation. \emph{J.~Phys.~Chem.~C} \textbf{2011}, \emph{115},
  16425--16432\relax
\mciteBstWouldAddEndPuncttrue
\mciteSetBstMidEndSepPunct{\mcitedefaultmidpunct}
{\mcitedefaultendpunct}{\mcitedefaultseppunct}\relax
\EndOfBibitem
\bibitem[Chaplais \latin{et~al.}(2018)Chaplais, Fraux, Paillaud, Marichal,
  Nouali, Fuchs, Coudert, and Patarin]{chap+18jpcc}
Chaplais,~G.; Fraux,~G.; Paillaud,~J.-L.; Marichal,~C.; Nouali,~H.;
  Fuchs,~A.~H.; Coudert,~F.-X.; Patarin,~J. Impacts of the Imidazolate Linker
  Substitution (CH3, Cl, or Br) on the Structural and Adsorptive Properties of
  ZIF-8. \emph{J.~Phys.~Chem.~C} \textbf{2018}, \emph{122}, 26945--26955\relax
\mciteBstWouldAddEndPuncttrue
\mciteSetBstMidEndSepPunct{\mcitedefaultmidpunct}
{\mcitedefaultendpunct}{\mcitedefaultseppunct}\relax
\EndOfBibitem
\bibitem[Yagi and Ueda(2023)Yagi, and Ueda]{yagi+23pccp}
Yagi,~R.; Ueda,~T. Substitution (CH 3, Cl, or Br) effects of the imidazolate
  linker on benzene adsorption kinetics for the zeolitic imidazolate framework
  (ZIF)-8. \emph{Phys.~Chem.~Chem.~Phys.~} \textbf{2023}, \emph{25},
  20585--20596\relax
\mciteBstWouldAddEndPuncttrue
\mciteSetBstMidEndSepPunct{\mcitedefaultmidpunct}
{\mcitedefaultendpunct}{\mcitedefaultseppunct}\relax
\EndOfBibitem
\bibitem[Amrouche \latin{et~al.}(2012)Amrouche, Creton, Siperstein, and
  Nieto-Draghi]{amro+12rsca}
Amrouche,~H.; Creton,~B.; Siperstein,~F.; Nieto-Draghi,~C. Prediction of
  thermodynamic properties of adsorbed gases in zeolitic imidazolate
  frameworks. \emph{RSC~Adv.} \textbf{2012}, \emph{2}, 6028--6035\relax
\mciteBstWouldAddEndPuncttrue
\mciteSetBstMidEndSepPunct{\mcitedefaultmidpunct}
{\mcitedefaultendpunct}{\mcitedefaultseppunct}\relax
\EndOfBibitem
\bibitem[D\"urholt \latin{et~al.}(2019)D\"urholt, Fraux, Coudert, and
  Schmid]{duer+19jctc}
D\"urholt,~J.~P.; Fraux,~G.; Coudert,~F.-X.; Schmid,~R. Ab Initio Derived Force
  Fields for Zeolitic Imidazolate Frameworks: MOF-FF for ZIFs.
  \emph{J.~Chem.~Theory.~Comput.~} \textbf{2019}, \emph{15}, 2420--2432\relax
\mciteBstWouldAddEndPuncttrue
\mciteSetBstMidEndSepPunct{\mcitedefaultmidpunct}
{\mcitedefaultendpunct}{\mcitedefaultseppunct}\relax
\EndOfBibitem
\bibitem[Morris \latin{et~al.}(2012)Morris, Stevens, Taylor, Dybowski, Yaghi,
  and Garcia-Garibay]{morris+12jpcc}
Morris,~W.; Stevens,~C.~J.; Taylor,~R.~E.; Dybowski,~C.; Yaghi,~O.~M.;
  Garcia-Garibay,~M.~A. NMR and X-ray Study Revealing the Rigidity of Zeolitic
  Imidazolate Frameworks. \emph{J.~Phys.~Chem.~C} \textbf{2012}, \emph{116},
  13307--13312\relax
\mciteBstWouldAddEndPuncttrue
\mciteSetBstMidEndSepPunct{\mcitedefaultmidpunct}
{\mcitedefaultendpunct}{\mcitedefaultseppunct}\relax
\EndOfBibitem
\bibitem[Ke \latin{et~al.}(2021)Ke, Duan, Ji, Zhao, Zhang, Duan, Li, and
  Wei]{Ke+21JCED}
Ke,~Q.; Duan,~Y.; Ji,~Y.; Zhao,~D.; Zhang,~H.; Duan,~C.; Li,~L.; Wei,~Y.
  Identical Composition and Distinct Performance: How ZIF-8 Polymorphs’
  Structures Affect the Adsorption/Separation of Ethane and Ethene. \emph{J.
  Chem. Eng. Data} \textbf{2021}, \emph{66}, 3483--3492\relax
\mciteBstWouldAddEndPuncttrue
\mciteSetBstMidEndSepPunct{\mcitedefaultmidpunct}
{\mcitedefaultendpunct}{\mcitedefaultseppunct}\relax
\EndOfBibitem
\bibitem[Hinuma \latin{et~al.}(2017)Hinuma, Pizzi, Kumagai, Oba, and
  Tanaka]{hinuma+17cms}
Hinuma,~Y.; Pizzi,~G.; Kumagai,~Y.; Oba,~F.; Tanaka,~I. Band structure diagram
  paths based on crystallography. \emph{Comp.~Mater.~Sci.~} \textbf{2017},
  \emph{128}, 140--184\relax
\mciteBstWouldAddEndPuncttrue
\mciteSetBstMidEndSepPunct{\mcitedefaultmidpunct}
{\mcitedefaultendpunct}{\mcitedefaultseppunct}\relax
\EndOfBibitem
\bibitem[Hohenberg and Kohn(1964)Hohenberg, and Kohn]{hohe-kohn64pr}
Hohenberg,~P.; Kohn,~W. Inhomogeneus Electron Gas. \emph{Phys.~Rev.~}
  \textbf{1964}, \emph{136}, B864--B871\relax
\mciteBstWouldAddEndPuncttrue
\mciteSetBstMidEndSepPunct{\mcitedefaultmidpunct}
{\mcitedefaultendpunct}{\mcitedefaultseppunct}\relax
\EndOfBibitem
\bibitem[Kohn and Sham(1965)Kohn, and Sham]{kohn-sham65pr}
Kohn,~W.; Sham,~L.~J. Self-Consistent Equations Including Exchange and
  Correlation Effects. \emph{Phys.~Rev.~} \textbf{1965}, \emph{140},
  A1133--A1138\relax
\mciteBstWouldAddEndPuncttrue
\mciteSetBstMidEndSepPunct{\mcitedefaultmidpunct}
{\mcitedefaultendpunct}{\mcitedefaultseppunct}\relax
\EndOfBibitem
\bibitem[Perdew \latin{et~al.}(1996)Perdew, Burke, and Ernzerhof]{pbe}
Perdew,~J.~P.; Burke,~K.; Ernzerhof,~M. Generalized Gradient Approximation Made
  Simple. \emph{Phys.~Rev.~Lett.~} \textbf{1996}, \emph{77}, 3865--3868\relax
\mciteBstWouldAddEndPuncttrue
\mciteSetBstMidEndSepPunct{\mcitedefaultmidpunct}
{\mcitedefaultendpunct}{\mcitedefaultseppunct}\relax
\EndOfBibitem
\bibitem[Giannozzi \latin{et~al.}(2020)Giannozzi, Baseggio, Bonfà, Brunato,
  Car, Carnimeo, Cavazzoni, de~Gironcoli, Delugas, Ferrari~Ruffino, Ferretti,
  Marzari, Timrov, Urru, and Baroni]{qe2020}
Giannozzi,~P.; Baseggio,~O.; Bonfà,~P.; Brunato,~D.; Car,~R.; Carnimeo,~I.;
  Cavazzoni,~C.; de~Gironcoli,~S.; Delugas,~P.; Ferrari~Ruffino,~F.
  \latin{et~al.}  Quantum ESPRESSO toward the exascale. \emph{J.~Chem.~Phys.~}
  \textbf{2020}, \emph{152}, 154105\relax
\mciteBstWouldAddEndPuncttrue
\mciteSetBstMidEndSepPunct{\mcitedefaultmidpunct}
{\mcitedefaultendpunct}{\mcitedefaultseppunct}\relax
\EndOfBibitem
\bibitem[Bl\"ochl(1994)]{bloechl94prb}
Bl\"ochl,~P.~E. Projector augmented-wave method. \emph{Phys.~Rev.~B}
  \textbf{1994}, \emph{50}, 17953--17979\relax
\mciteBstWouldAddEndPuncttrue
\mciteSetBstMidEndSepPunct{\mcitedefaultmidpunct}
{\mcitedefaultendpunct}{\mcitedefaultseppunct}\relax
\EndOfBibitem
\bibitem[{Dal Corso}(2014)]{DALCORSO14cms}
{Dal Corso},~A. Pseudopotentials periodic table: From H to Pu.
  \emph{Comp.~Mater.~Sci.~} \textbf{2014}, \emph{95}, 337--350\relax
\mciteBstWouldAddEndPuncttrue
\mciteSetBstMidEndSepPunct{\mcitedefaultmidpunct}
{\mcitedefaultendpunct}{\mcitedefaultseppunct}\relax
\EndOfBibitem
\bibitem[Bader and Zou(1992)Bader, and Zou]{BADER92cpl}
Bader,~R.; Zou,~P. An atomic population as the expectation value of a quantum
  observable. \emph{Chem.~Phys.~Lett.~} \textbf{1992}, \emph{191}, 54--58\relax
\mciteBstWouldAddEndPuncttrue
\mciteSetBstMidEndSepPunct{\mcitedefaultmidpunct}
{\mcitedefaultendpunct}{\mcitedefaultseppunct}\relax
\EndOfBibitem
\bibitem[de-la Roza \latin{et~al.}(2014)de-la Roza, Johnson, and
  Luaña]{oter+14cpc}
de-la Roza,~A.~O.; Johnson,~E.~R.; Luaña,~V. Critic2: A program for real-space
  analysis of quantum chemical interactions in solids.
  \emph{Comput.~Phys.~Commun.~} \textbf{2014}, \emph{185}, 1007--1018\relax
\mciteBstWouldAddEndPuncttrue
\mciteSetBstMidEndSepPunct{\mcitedefaultmidpunct}
{\mcitedefaultendpunct}{\mcitedefaultseppunct}\relax
\EndOfBibitem
\bibitem[Yu and Trinkle(2011)Yu, and Trinkle]{yu+trin11jcp}
Yu,~M.; Trinkle,~D.~R. Accurate and efficient algorithm for Bader charge
  integration. \emph{J.~Chem.~Phys.~} \textbf{2011}, \emph{134}, 064111\relax
\mciteBstWouldAddEndPuncttrue
\mciteSetBstMidEndSepPunct{\mcitedefaultmidpunct}
{\mcitedefaultendpunct}{\mcitedefaultseppunct}\relax
\EndOfBibitem
\bibitem[Guilherme~Buzanich \latin{et~al.}(2023)Guilherme~Buzanich, Radtke,
  Yusenko, M.~Stawski, Kulow, Cakir, Roeder, Naese, Britzke, Sintschuk, and
  Emmerling]{guil+23jcp}
Guilherme~Buzanich,~A.; Radtke,~M.; Yusenko,~K.~V.; M.~Stawski,~T.; Kulow,~A.;
  Cakir,~C.~T.; Roeder,~B.; Naese,~C.; Britzke,~R.; Sintschuk,~M.
  \latin{et~al.}  {BAMline—A real-life sample materials research beamline}.
  \emph{J.~Chem.~Phys.~} \textbf{2023}, \emph{158}, 244202\relax
\mciteBstWouldAddEndPuncttrue
\mciteSetBstMidEndSepPunct{\mcitedefaultmidpunct}
{\mcitedefaultendpunct}{\mcitedefaultseppunct}\relax
\EndOfBibitem
\bibitem[Ravel and Newville(2005)Ravel, and Newville]{rave-newv05jsr}
Ravel,~B.; Newville,~M. ATHENA, ARTEMIS, HEPHAESTUS: data analysis for X-ray
  absorption spectroscopy using IFEFFIT. \emph{J.~Synchrotron~Radiat.}
  \textbf{2005}, \emph{12}, 537--541\relax
\mciteBstWouldAddEndPuncttrue
\mciteSetBstMidEndSepPunct{\mcitedefaultmidpunct}
{\mcitedefaultendpunct}{\mcitedefaultseppunct}\relax
\EndOfBibitem
\bibitem[Cui and Schmidt(2020)Cui, and Schmidt]{cui-schm20jpcc}
Cui,~K.; Schmidt,~J. Enabling efficient and accurate computational studies of
  MOF reactivity via QM/MM and QM/QM methods. \emph{J.~Phys.~Chem.~C}
  \textbf{2020}, \emph{124}, 10550--10560\relax
\mciteBstWouldAddEndPuncttrue
\mciteSetBstMidEndSepPunct{\mcitedefaultmidpunct}
{\mcitedefaultendpunct}{\mcitedefaultseppunct}\relax
\EndOfBibitem
\bibitem[Hartmann \latin{et~al.}(1997)Hartmann, Clark, and van
  Eldik]{hart+97jacs}
Hartmann,~M.; Clark,~T.; van Eldik,~R. Hydration and water exchange of zinc
  (II) ions. Application of density functional theory.
  \emph{J.~Am.~Chem.~Soc.~} \textbf{1997}, \emph{119}, 7843--7850\relax
\mciteBstWouldAddEndPuncttrue
\mciteSetBstMidEndSepPunct{\mcitedefaultmidpunct}
{\mcitedefaultendpunct}{\mcitedefaultseppunct}\relax
\EndOfBibitem
\bibitem[Bock \latin{et~al.}(1999)Bock, Katz, Markham, and
  Glusker]{bock+99jacs}
Bock,~C.~W.; Katz,~A.~K.; Markham,~G.~D.; Glusker,~J.~P. Manganese as a
  replacement for magnesium and zinc: functional comparison of the divalent
  ions. \emph{J.~Am.~Chem.~Soc.~} \textbf{1999}, \emph{121}, 7360--7372\relax
\mciteBstWouldAddEndPuncttrue
\mciteSetBstMidEndSepPunct{\mcitedefaultmidpunct}
{\mcitedefaultendpunct}{\mcitedefaultseppunct}\relax
\EndOfBibitem
\bibitem[Sandmark and Br{\"a}nden(1967)Sandmark, and Br{\"a}nden]{sand-brae67}
Sandmark,~C.; Br{\"a}nden,~C.-I. The Crystal Structure of Hexaimidazole Zinc
  (II) Dichloride Tetrahydrate, Zn(C$_3$H$_4$N$_2$)$_6$Cl2$\cdot$4H$_2$O.
  \emph{Acta~Chem.~Scand.} \textbf{1967}, \emph{21}\relax
\mciteBstWouldAddEndPuncttrue
\mciteSetBstMidEndSepPunct{\mcitedefaultmidpunct}
{\mcitedefaultendpunct}{\mcitedefaultseppunct}\relax
\EndOfBibitem
\bibitem[Torzilli \latin{et~al.}(2002)Torzilli, Colquhoun, Doucet, and
  Beer]{torz+02polyhedron}
Torzilli,~M.~A.; Colquhoun,~S.; Doucet,~D.; Beer,~R.~H. The interconversion of
  dichlorobis (Nn-propylsalicylaldimine) zinc (II) and bis
  (Nn-propylsalicylaldiminato) zinc (II). \emph{Polyhedron} \textbf{2002},
  \emph{21}, 697--704\relax
\mciteBstWouldAddEndPuncttrue
\mciteSetBstMidEndSepPunct{\mcitedefaultmidpunct}
{\mcitedefaultendpunct}{\mcitedefaultseppunct}\relax
\EndOfBibitem
\bibitem[Boada \latin{et~al.}(2022)Boada, Chaboy, Hayama, Keenan, Freeman,
  Amboage, and D{\'\i}az-Moreno]{boad+22jpcc}
Boada,~R.; Chaboy,~J.; Hayama,~S.; Keenan,~L.~L.; Freeman,~A.~A.; Amboage,~M.;
  D{\'\i}az-Moreno,~S. Unraveling the Molecular Details of the “Gate
  Opening” Phenomenon in Zif-8 with X-Ray Absorption Spectroscopy.
  \emph{J.~Phys.~Chem.~C} \textbf{2022}, \emph{126}, 5935--5943\relax
\mciteBstWouldAddEndPuncttrue
\mciteSetBstMidEndSepPunct{\mcitedefaultmidpunct}
{\mcitedefaultendpunct}{\mcitedefaultseppunct}\relax
\EndOfBibitem
\bibitem[Cocchi \latin{et~al.}(2011)Cocchi, Prezzi, Ruini, Caldas, and
  Molinari]{cocc+11jpcl}
Cocchi,~C.; Prezzi,~D.; Ruini,~A.; Caldas,~M.~J.; Molinari,~E. Optical
  properties and charge-transfer excitations in edge-functionalized
  all-graphene nanojunctions. \emph{J.~Phys.~Chem.~Lett.} \textbf{2011},
  \emph{2}, 1315--1319\relax
\mciteBstWouldAddEndPuncttrue
\mciteSetBstMidEndSepPunct{\mcitedefaultmidpunct}
{\mcitedefaultendpunct}{\mcitedefaultseppunct}\relax
\EndOfBibitem
\end{mcitethebibliography}

\providecommand{\latin}[1]{#1}
\makeatletter
\providecommand{\doi}
  {\begingroup\let\do\@makeother\dospecials
  \catcode`\{=1 \catcode`\}=2 \doi@aux}
\providecommand{\doi@aux}[1]{\endgroup\texttt{#1}}
\makeatother
\providecommand*\mcitethebibliography{\thebibliography}
\csname @ifundefined\endcsname{endmcitethebibliography}
  {\let\endmcitethebibliography\endthebibliography}{}

\end{document}